\documentclass[aps,reprint,pra,longbibliography,groupedaddress]{revtex4-2}
\usepackage{}
\usepackage{amssymb}
\usepackage{amsmath}
\usepackage{dcolumn}
\usepackage{bm}
\usepackage{graphicx}
\usepackage{mathrsfs}
\usepackage[colorlinks,linkcolor=blue,anchorcolor=blue,citecolor=blue,urlcolor=blue]{hyperref}
\usepackage{color}
\UseRawInputEncoding

\begin{document}

\preprint{APS/123-QED}

\title{Nonreciprocal phonon blockade}

\author{Xiao-Yu Yao}
\affiliation{Ministry of Education Key Laboratory for Nonequilibrium Synthesis and Modulation of Condensed Matter, Shaanxi Province Key Laboratory of Quantum Information and Quantum Optoelectronic Devices, School of Physics, Xi'an Jiaotong University, Xi'an 710049, China }

\author{Hamad Ali}
\affiliation{Ministry of Education Key Laboratory for Nonequilibrium Synthesis and Modulation of Condensed Matter, Shaanxi Province Key Laboratory of Quantum Information and Quantum Optoelectronic Devices, School of Physics, Xi'an Jiaotong University, Xi'an 710049, China }

\author{Peng-Bo Li}
 \email{lipengbo@mail.xjtu.edu.cn}
\affiliation{Ministry of Education Key Laboratory for Nonequilibrium Synthesis and Modulation of Condensed Matter, Shaanxi Province Key Laboratory of Quantum Information and Quantum Optoelectronic Devices, School of Physics, Xi'an Jiaotong University, Xi'an 710049, China }

\date{\today}

\begin{abstract}
Quantum nonreciprocal devices have received extensive attention in recent years because they can be used to realize unidirectional quantum routing and noise isolation. In this work, we show that the shift of resonance frequencies of propagating phonons induced by spin-orbit interactions (SOI) of phonons in a rotating acoustic ring cavity can be used to realize nonreciprocal phonon blockade. When driving the cavity from different directions, nonreciprocal single-, two-phonon blockade and phonon-induced tunneling can take place by varying the parameters of the system to an appropriate value. To realize phonon blockade, the sublevels of the lower orbit branch of the ground state of silicon-vacancy (SiV) color centers in the diamond membrane are employed to induce self-interactions of phonons in the cavity. This work provides a way to achieve acoustic nonreciprocal devices, such as directional acoustic switches and quantum noise isolation, which may help acoustic information network processing.

\end{abstract}

\maketitle


\section{introduction}
In quantum information processing, reliable methods for creation and manipulation of phonons have become the subject of interests on account of their potential applications~\cite{1,2,3,4,5}. Phononic devices work in the GHz frequency range that corresponds to wavelengths on the order of $\mu$m, which makes them a bridge between optical and artificial circuits~\cite{6,7}. Thus it has the potential to be a good candidate for solid-state quantum devices~\cite{8,9,10,11,12}. Various methods and architectures have been investigated to study quantum phononic effects, such as phonon blockade (PB).

Phonon blockade is a phenomenon when one phonon is present in a nonlinear acoustic cavity, the subsequent phonons cannot be excited, which is one of the important mechanisms to realize the single phonon source~\cite{13}. In close analogy to  Coulomb blockade~\cite{14} and photon blockade~\cite{15,16,17,18,19,20,21,22},  PB is a pure quantum effect, which plays an important role in unveiling the quantum behavior of a device. The initial single-phonon blockade (1PB) was achieved in a nanomechanical resonator, which is coupled to a superconducting quantum two-level system for inducing self-interactions of phonons. On this basis, great effort has been made to study other effects such as two-phonon blockade (2PB), equivalent phonon-induced tunneling (PIT), and detecting of PB ~\cite{23,24,25,26,27,28}.
Optomechanical systems with second order non-linearities are considered for photon induced phonon blockade~\cite{23,29,31}, with potential applications in hybrid photon-phonon quantum networks.

Recently, nonreciprocal devices have received widespread attention~\cite{36,37,38,39,40}. Reciprocity is a basic theorem in a network, which states that when the source and observation points of the network change the positions, the response of the transmission channel is symmetric~\cite{36,41}.  Nonreciprocal devices, allowing the flow of signal from one side but blocking it from the other, have become fundamental devices for building information processing networks, such as unidirectional transmission, cloaking and noise-free information processing and amplifiers~ \cite{36,42,43,44,45,46,47,48}. In classical electromagnetic networks, one usually uses magnetic materials or strong nonlinearities to destroy reciprocity ~\cite{49,50,51,52,54}. But in the quantum networks, magnetic materials are bulky and hard to integrate with other devices. Furthermore, by using nonlinear materials, to achieve nonreciprocity requires higher input strength from only one port at a time. So in recent years, using temporal modulation to break reciprocity has received much attention, such as traveling wave modulation ~\cite{55,56,57,58,59,60,61}, Sagnac effect of light ~\cite{62,63,64,65}, and direct photon transition  ~\cite{66,67,68,69,70}, which have great potential for integrated nonreciprocal devices. For acoustic networks, nonreciprocal devices have also made great progress ~\cite{71,72,73,74,75,76,77,78,79,80,81,82}. It is reported that the use circulating fluids and synthetic angular momentum in acoustic meta-atom or meta-materials can realize acoustic circulators~\cite{71}. On this basis, many other works have been performed, such as using acoustic metamaterials and sonic crystal structures to achieve unidirectional acoustic transmission ~\cite{73,74,75}, and nonreciprocal phonon lasers based on optomechanical systems~\cite{78}. But so far, the nonreciprocal realization of phonon blockade has yet to realize.

In this work we propose a spin-phononic system to study nonreciprocal PB and PIT in non-stationary ring resonators, having counter prorogating travelling acoustic modes coupled to color centers in diamond.  The spin-phononic system~\cite{85}, which can be regarded as a phonon cavity quantum electrodynamics(QED) system, may provide a promising setup for realizing nonreciprocal PB. The mechanical vibrations can be controlled through their couplings with spins~\cite{86,87,83,84}. And NV and SiV color centers in diamond have long become promising qubit platforms~\cite{88,89,90,91,92}. Especially, the SiV has relatively strong strain coupling, which can be used to induce phonon-phonon interactions~\cite{93}, and has good potential for achieving phonon blockade.

To be specific, we use the sublevels of the lower orbit branch of the ground state of the SiV color centers as a two-level system to induce phonon self-interactions in the rotating ring cavity~\cite{94,95}. We find that, nonreciprocal 1PB and PIT can be induced in the spinning ring resonator when driving it from the left- or right-side of a coupled phononic waveguide, because the cavity phonons satisfy the sub-Poisson distribution or the super-Poisson distribution. Furthermore, by varying other parameters of the system, we can get other nonreciprocal phenomena such as 2PB and multi-PB. The physical origins behind the nonreciprocity phenomena are the SOI of phonons in the spinning ring resonator.  It cause a shift in the resonance frequency and phase of the propagating phonon mode in the cavity, which has been proved in recent experiments ~\cite{96}. Such nonreciprocal PB can be used for many acoustic devices, such as acoustic diodes~\cite{97,98,99,100,101}, which can play a key role in a phonon-basis information network at the few-phonon levels~\cite{102}.

The remainder of this paper is organized as follows. In Sec. II, we describe the physical model of a acoustic ring resonator coupled to a two-level SiV center in diamond. In Sec. III, we present phonon bloackade in case of the stationary ring resonator coupled to two level system. In Sec. IV, we present our result in case of the rotating system. Experimental feasibility of our scheme is given in Sec. V. Finally, the conclusion is given in Sec. VI.
\section{Model}
\begin{figure}[t]
\includegraphics[width=8cm]{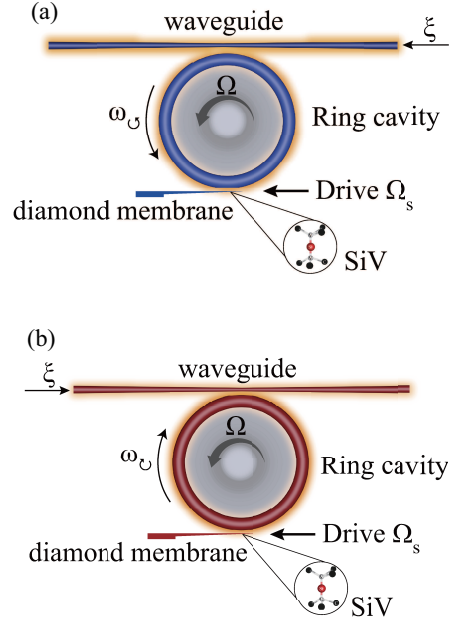}
\caption{\label{fig:1}(Color online)
 Schematic of the composite spin-phononic system. One side of the ring cavity is coupled with the phononic waveguide to the input and output of the driving field, and the other side is coupled with the SiV centers in the diamond membrane to realize phonon blockade. In addition, we add a rotator to the cavity to make the cavity rotate counterclockwise at an angular frequency $\Omega$. The phonons which input from the left- or right-side of the waveguide rotates counterclockwise or clockwise in the cavity corresponding to (a) or (b), respectively.}
\end{figure}

The system we considered consists of an acoustic ring cavity coupled to  a phononic waveguide on one side and to SiV centers in diamond membrane on the other side, as illustrated in Fig.~\ref{fig:1}.
The cavity and the diamond membrane are contact-coupled~\cite{103,104,105}, and the sublevels of the lower orbital branch of ground-state of SiV centers are coupled to the phonon mode in the cavity. Under the rotating wave approximation and neglecting two-phonon terms, the Hamiltonian of such system is given by
\begin{eqnarray}\label{ME1}
H^{\text {(0)}}&=&\hbar \omega a^{\dagger} a+\frac{1}{2} \hbar \omega_{0} \sigma_{z}+\hbar g\left(a \sigma_{+}+a^{\dagger} \sigma_{-}\right))\notag\\
&+&\hbar \Omega_{s}\left(\sigma_{+} e^{-i \omega_{1} t}+\sigma_{-} e^{i \omega_{1} t}\right),\notag\\
H^{\text {(d)}}&=&\hbar\xi\left(a^{\dagger}e^{-i \omega_{L} t}+ae^{i \omega_{L} t} \right),
\end{eqnarray}
here, $\omega$ is the resonance frequency of the cavity, $a$$(a^{\dagger})$ denote the annihilation(creation) operator of the cavity field, $\hbar\omega_{0}$ is the level splitting of the two-level system, and SiV centers is descried by the spin operator $ \sigma_{z}=|e\rangle\langle e|-|g \rangle\langle g|$, with $|e\rangle$ and $|g\rangle$ represent the excited and ground state of SiV centers, respectively. The ladder operators $\sigma^{\pm}=\frac{1}{2}\left(\sigma^{x} \pm i \sigma^{y}\right)$ describe the interaction between the SiV centers and the phononic cavity, with $ \sigma^{x}$ and $\sigma^{y}$ as Pauli matrices, where $g$ is the coupling strength between the cavity mode and SiV center. The parameter $\xi$ is the driving field amplitude with driving frequency $\omega_{L}$.

In the regime where SiV centers and phonon modes are far detuned to the coupling strength $
\Delta=\left(\omega-\omega_{0}\right) \gg g$ and the Rabi frequency $\Omega_s$, by which SiV centers are driven with the field of frequency $\omega_{1}$ obeys the condition $\Omega_{s} \gg \left(g^2/\Delta \right)$. Here we assume that $\omega_1=\omega_0$, and then the Hamiltonian $H^{\text {(0)}}$ in Eq.(\ref{ME1}) in the rotating frame with $V= \exp (-i\omega_0\sigma_z t/2)$  can be written as~\cite{13,106}
\begin{eqnarray}\label{ME2}
H^{\text {(0)}}_{\text {eff}}= \hbar \omega a^{\dagger} a+ \hbar Ua^{\dagger}a^{\dagger}aa,
\end{eqnarray}
with the effective phonon-phonon interaction term (nonlinearity strength)
\begin{eqnarray}\label{ME3}
U=\frac{g^{4}}{\Omega_{s}\Delta^{2}}.
\end{eqnarray}
Here, it is worth mentioning that after straightforward calculations we have ignored $ \rho_{z}=|+\rangle\langle+|-|-\rangle\langle-|$, with $	|\pm\rangle=(|g\rangle \pm|e\rangle) / \sqrt{2} $, by preparing the SiV in the ground state $|-\rangle$.

In the ring cavity, the phonon modes with two different propagation directions  (clockwise (CW) or counter-clockwise (CCW), corresponding to driving of the cavity from the left- or right-hand, respectively) carry different orbital angular momentum (OAM)~\cite{96,107}. The phonon in different propagation directions will possess spins of opposite direction because of SOI. Therefore, phonons propagating CW or CCW experience the centripetal or centrifugal Coriolis force, respectively \cite{108}. When we add a rotator that rotates CCW at the angular frequency $\Omega$ to the cavity, the resonant frequency of phonons will have a change, i.e. $\omega \rightarrow \omega +\Delta_{F}$, where
\begin{eqnarray}\label{ME4}
\Delta_{F}=\chi\Omega,
\end{eqnarray}
here $\chi=\pm0.12$ is the chirality of the phonon. For the rotator, rotating in the CCW direction, when the cavity phonon rotates CW, $\Delta_{F}>0$, otherwise, $\Delta_{F}<0$. Now in the rotating reference frame at the driving frequency $\omega_{L}$, the effective Hamiltonian of the driven  system can be written as:
\begin{eqnarray}\label{ME5}
H_{\text {eff}}=& \hbar&\left(-\Delta_{L}+\Delta_{F}\right)  a^{\dagger} a+\hbar U a^{\dagger}a^{\dagger}aa\notag\\
&+&\hbar\xi\left(a^{\dagger}+a \right),
\end{eqnarray}
here, $\Delta_{L}=\omega_{L}-\omega$ is the detuning between the field of the non-rotation resonant cavity and the driving field.

In order to obtain the analytical solution of the system, we first consider the isolated phononic system. We have
\begin{eqnarray}\label{ME6}
H_{\text {eff}}=\hbar\left(-\Delta_{L}+\Delta_{F}\right)  a^{\dagger} a+\hbar U a^{\dagger}a^{\dagger}aa.
\end{eqnarray}
For a Fock state $\vert n\rangle$, we have
\begin{eqnarray*}\nonumber
	H_{\text {eff}}|n\rangle &=&\left[-\hbar \Delta_{L} \hat{a}^{\dagger} \hat{a}+\hbar \Delta_{F} \hat{a}^{\dagger} \hat{a}+\hbar U a^{\dagger}a^{\dagger}aa\right]|n\rangle \\
	&=&\left[-\hbar \Delta_{L} \hat{a}^{\dagger} \hat{a}+\hbar\left( \Delta_{F}-U \right) \hat{a}^{\dagger} \hat{a}+\hbar U \hat{a}^{\dagger} \hat{a} \hat{a}^{\dagger} \hat{a}\right]|n\rangle \\
	&=&\left[-n \hbar \Delta_{L}+n \hbar\left(\Delta_{F}-U\right)+n^{2} \hbar U\right]|n\rangle \\
	&=&E_{n}|n\rangle.
\end{eqnarray*}
Thus we obtain the eigenenergy
\begin{eqnarray}\label{ME7}
E_{n}=-n\hbar \Delta_{L}+n \hbar \Delta_{F}+\left(n^{2}-n\right) \hbar U,
\end{eqnarray}
where n is the number of cavity phonons. The eigenenergy is also effective for the weak driving condition. In the region of weak drive $\xi \ll \gamma$, we can truncate the Hilbert space of the system to n=2. Here, $\gamma=\omega/Q$ is the cavity dissipation rate and $Q$ is the quality factor. The state of the system can be written as $|\varphi(t)\rangle=\sum_{n=0}^{2} C_{n}(t)|n\rangle$, with $ C_{n}$ is the probability amplitude.  For the effective Hamiltonian (Eq.5), we can introduce a dissipation term  according to the quantum-trajectory method~\cite{109}:   $H=H_{\text {eff}}-(i \hbar \gamma / 2) \hat{a}^{\dagger} \hat{a}$. Then the equation of motion of the dissipative  system is given by
\begin{eqnarray}\label{ME8}
	&\dot{C}_{0}(t)&=-i \frac{E_{0}}{\hbar} C_{0}(t)-i \xi C_{1}(t) ,\notag\\
	&\dot{C}_{1}(t)&=-i\left(\frac{E_{1}}{\hbar}-i \frac{\gamma}{2}\right) C_{1}(t)-i \xi C_{0}(t)-i \xi \sqrt{2} C_{2}(t), \notag \\
	&\dot{C}_{2}(t)&=-i\left(\frac{E_{2}}{\hbar}-i \gamma\right) C_{2}(t)-i \xi \sqrt{2} C_{1}(t).
\end{eqnarray}
With the initial conditions: $C_{0}(0)=1$, and $C_{1}(0)=C_{2}(0)=0$, the Steady-state solutions of the system can be derived as
\begin{eqnarray}\label{ME9}
&C_{1}(\infty)&=\frac{-\xi}{\left(\frac{E_{1}}{\hbar}-\frac{E_{0}}{\hbar}-i \frac{\gamma}{2}\right)}, \notag\\
&C_{2}(\infty)&=\frac{-\sqrt{2} \xi C_{1}(\infty)}{\left(\frac{E_{2}}{\hbar}-\frac{E_{0}}{\hbar}-i \gamma\right)}.
\end{eqnarray}

Now we proceed to study PB and PIT. For this purpose, we will exploit the $
\mu^{th}$-order correlation function of zero-time delay $g^{(\mu)}(0) \equiv\left\langle\hat{a}^{\dagger \mu} \hat{a}^{\mu}\right\rangle /\langle\hat{n}\rangle^{\mu}$, with $\hat{n}=\hat{a}^{\dagger}\hat{a}$.
Describing the probability of finding n phonons in the acoustic cavity by $P(n)=|C(n)|^{2}$, we can define the correlation function as follows
\begin{eqnarray}\label{ME10}
g^{(2)}(0)&=&\frac{2 P_{2}}{\left(P_{1}+2 P_{2}\right)^{2}} \notag\\
&\simeq& \frac{\left(-\Delta_{L}+\Delta_{F}\right)^{2}+\gamma^{2} / 4}{\left(-\Delta_{L}+\Delta_{F}+U\right)^{2}+\gamma^{2} / 4}.
\end{eqnarray}
For n=3, we can get the third-order correlation function
\begin{eqnarray}\label{ME11}
&&g^{(3)}(0)=\frac{6 P_{3}}{\left(P_{1}+2 P_{2}+3 P_{3}\right)^{3}} \notag\\
&\simeq&\frac{\left[\left(\Delta_{L}+\Delta_{F}\right)^{2}+\frac{\gamma^{2}}{4}\right]^{2}}{\left[\left(\Delta_{L}+\Delta_{F}+U\right)^{2}+\frac{\gamma^{2}}{4}\right]\left[\left(\Delta_{L}+\Delta_{F}+2 U\right)^{2}+\frac{\gamma^{2}}{4}\right]}.\notag\\
\end{eqnarray}

The minimum and the maximum of $g^{(\mu)}(0)$ are corresponding to 1PB, multi-PB, and PIT, respectively. For example, when $g_{min}^{(2)}(0)=1/[4\left(U/\gamma\right)^{2}+1]  \textless  1$ for $\Delta_{L}=\Delta_{F}$, the phonons satisfy the super-Poisson distribution (phonon bunching state) which corresponds to 1PB; when  $g_{max}^{(2)}(0)=4\left(U/\gamma\right)^{2}+1 \textgreater 1$ for $\Delta_{L}= \Delta_{F}+U$, the phonons satisfy the sub-Poisson distribution (phonon anti-bunching state), which corresponds to PIT. For $\mu\textgreater2$, we have similar results for multi-PB and PIT.

\section{Phonon blockade}

In this section, we study the realization of PB in the hybrid quantum system including the dissipation effect. The master equation for the system is given as,

\begin{eqnarray}\label{ME12}
\dot{\hat{\rho}}=\frac{i}{\hbar}[\hat{\rho}, H_{\text {eff}}]+\frac{\gamma}{2}\left(2 \hat{a} \hat{\rho} \hat{a}^{\dagger}-\hat{a}^{\dagger} \hat{a} \hat{\rho}-\hat{\rho} \hat{a}^{\dagger} \hat{a}\right),
\end{eqnarray}
where $H_{\text {eff}}$ is the hamiltonian given in Eq. 5. We can get the phonon number probability $P(n)=\left\langle n\left|\hat{\rho}_{\mathrm{}}\right| n\right\rangle$ from the steady-state solution $\hat{\rho}_{\mathrm{}}$ of the system.

\begin{figure}[b]
	\includegraphics[width=8.6cm]{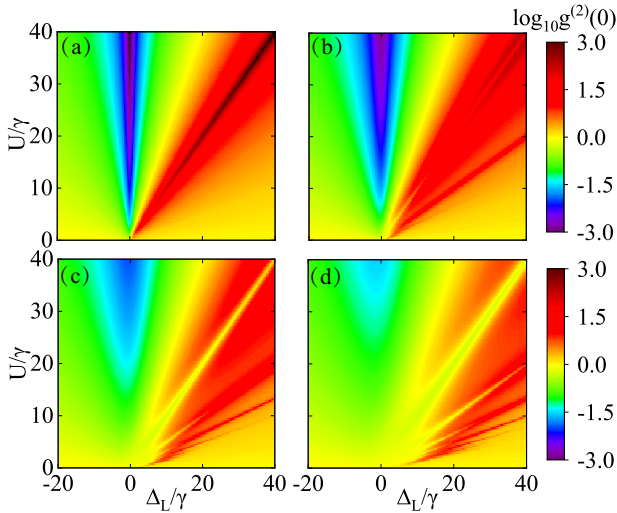}
	\caption{\label{fig:2}(Color online)
		$g^{(2)}(0)$ as a function of the nonlinear strength $U$ and the detuning $\Delta_{L}$, here $\Delta_{F}=0$. (a), (b), (c), and (d) corresponds to the situation when $\xi=0.33\gamma$,  $\xi=1\gamma$, $\xi=3\gamma$, and $\xi=5\gamma$ respectively.}
\end{figure}

We now focus on the properties of the cavity phonons described by the second order correlation function for various deriving field strengths $\xi$. In Fig.~\ref{fig:2} we plot the results given by Eq.~(\ref{ME12}) as a function of the non-linear strength U and detuning $\Delta_L$ in the case of a stationary cavity i.e. $\Delta_{F}=0$. In the weak drive condition ($\xi=0.33\gamma$), one can easily see the maximum and minimum regions of the second-order correlation function shown in Fig.~\ref{fig:2}(a). The multiple relative maximum curves in Fig.~\ref{fig:2}(b), (c) and~(d) corresponding to $\xi=1\gamma$, $\xi=3\gamma$ and $\xi=5\gamma$ respectively, are derived from the contribution of multi-phonon resonances rather than just two-phonon resonance; this effect does not appear in the weak drive case. At this time, the entire system is equivalent to a conventional PB model. Here we introduce a standard criterion for judging PB and PIT  \cite{110,111,112}.

For n-phonon blockade (nPB), there is
\begin{eqnarray}\label{ME13}
&(&i) \quad g^{(n+1)}(0)<\exp (-\langle\hat{m}\rangle) = f,\quad \notag\\
&(&ii) \quad g^{(n)}(0) \geq \exp (-\langle\hat{m}\rangle)+\langle\hat{m}\rangle \cdot g^{(n+1)}(0) = f^{(n)};
\end{eqnarray}
for PIT, there is
\begin{eqnarray}\label{ME14}
g^{(\mu)}(0)>\exp (-\langle\hat{m}\rangle), \quad \mathrm{for}\quad \mu\geq2.
\end{eqnarray}
Here, $\langle\hat{m}\rangle$ is the average phonon number of the cavity field.

In this work, we mainly consider the situation where the number of phonons is small (n$\leq$4). In this case, $\langle\hat{m}\rangle\ll 1$ , so Eq. (\ref{ME14}) can be rewritten as
\begin{eqnarray}\label{ME15}
	g^{(\mu)}(0)>1, \quad \mathrm{for}\quad 4\geq \mu\geq 2.
\end{eqnarray}

For nPB, we also have an equivalent standard for comparing the phonon number distribution with the Poisson distribution:
\begin{eqnarray}\label{ME16}
&(&i) \quad P(m)<\mathcal{P}(m), \quad \mathrm{for}\quad  m>n\notag\\
&(&ii)\quad P(n) \geq \mathcal{P}(n),
\end{eqnarray}
with $\mathcal{P}(n)=\frac{\langle\hat{n}\rangle^{n}}{n !} \exp (-\langle\hat{n}\rangle)$ is the Poissonian distribution.
\begin{figure}[b]
\includegraphics[width=8.6cm]{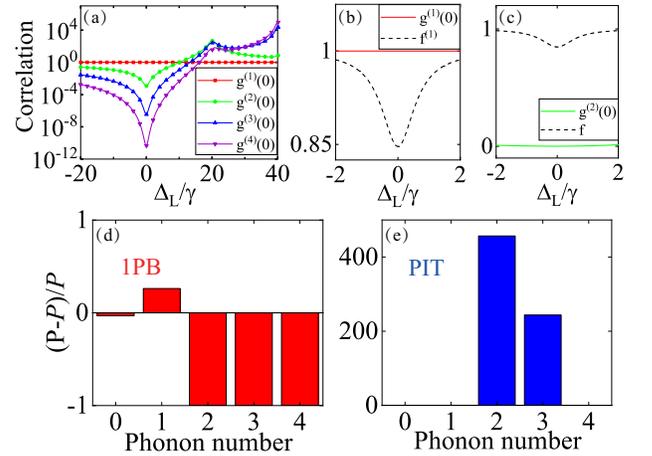}
\caption{\label{fig:3}(Color online)
(a) Under the weak driving strength region $\xi=0.33\gamma$, the correlation function versus the detuning $\Delta_{L}$ at $U=20\gamma$. (b) and (c) show the comparison for 1PB between the correlation function and the standard value in Eq. (\ref{ME18}). (d) and (e) show the deviation between the phonon population and the Poisson distribution for 1PB and PIT, respectively at $\Delta_{L}=0$ and $\Delta_{L}=20\gamma$.}
\end{figure}
To show a relative deviation of a given phonon-number distribution from the corresponding Poissonian distribution, we use the formula \cite{112}
\begin{eqnarray}\label{ME17}
[P(n)-\mathcal{P}(n)] / \mathcal{P}(n).
\end{eqnarray}

To further elaborate the effect of weak drive on PB in the non-rotating system ($\Delta_{\mathrm {F}}= 0$), in Fig.~\ref{fig:3} we have plotted the n-order correlation function for constant non-linear strength (U). Figure 3(a) shows the results for the n-order correlation function with respect to the detuning $(\Delta_L)$ for $U=20\gamma$. Here, for $g^{(2)}(0)$,  the minimum and maximum values are at $\Delta_{L}=0$ and $\Delta_{L}=20\gamma$ , which corresponds to 1PB and PIT respectively.

According to Eq. (\ref{ME13}), 1PB should meet the following conditions for $n=1$:
\begin{eqnarray}\label{ME18}
	&(&i) \quad g^{(2)}(0)<\exp (-\langle\hat{m}\rangle) = f,\notag\\
	&(&ii)\quad g^{(1)}(0) \geq \exp (-\langle\hat{m}\rangle)+\langle\hat{m}\rangle \cdot g^{(2)}(0) = f^{(1)}.
\end{eqnarray}
Next, in Fig.~\ref{fig:3}(b) and Fig.~\ref{fig:3}(c) we plot the n-order correlation functions and standard values given in Eq.~18 as functions of $\Delta_L$. We find that the curves of the correlation function satisfy the conditions in Eq.(\ref{ME18}). Furthermore, in Fig.~\ref{fig:3}(d), we find that $P(1)>\mathcal{P}(1)$, $P(2)<\mathcal{P}(2)$, and $P(3)<\mathcal{P}(3)$, which is a clear signature of 1PB. Due to PIT corresponding to the super-Poisson distribution of phonons, we can find that $P(2)>\mathcal{P}(2)$ and $P(3)>\mathcal{P}(3)$ in Fig.~\ref{fig:3}(e). In addition, we can find that $g^{(2)}(0)>g^{(3)}(0)>g^{(4)}(0)>1$ in Fig.~\ref{fig:3}(a), which indicates the emergence of PIT at this time.

\begin{figure}[b]
\includegraphics[width=8.6cm]{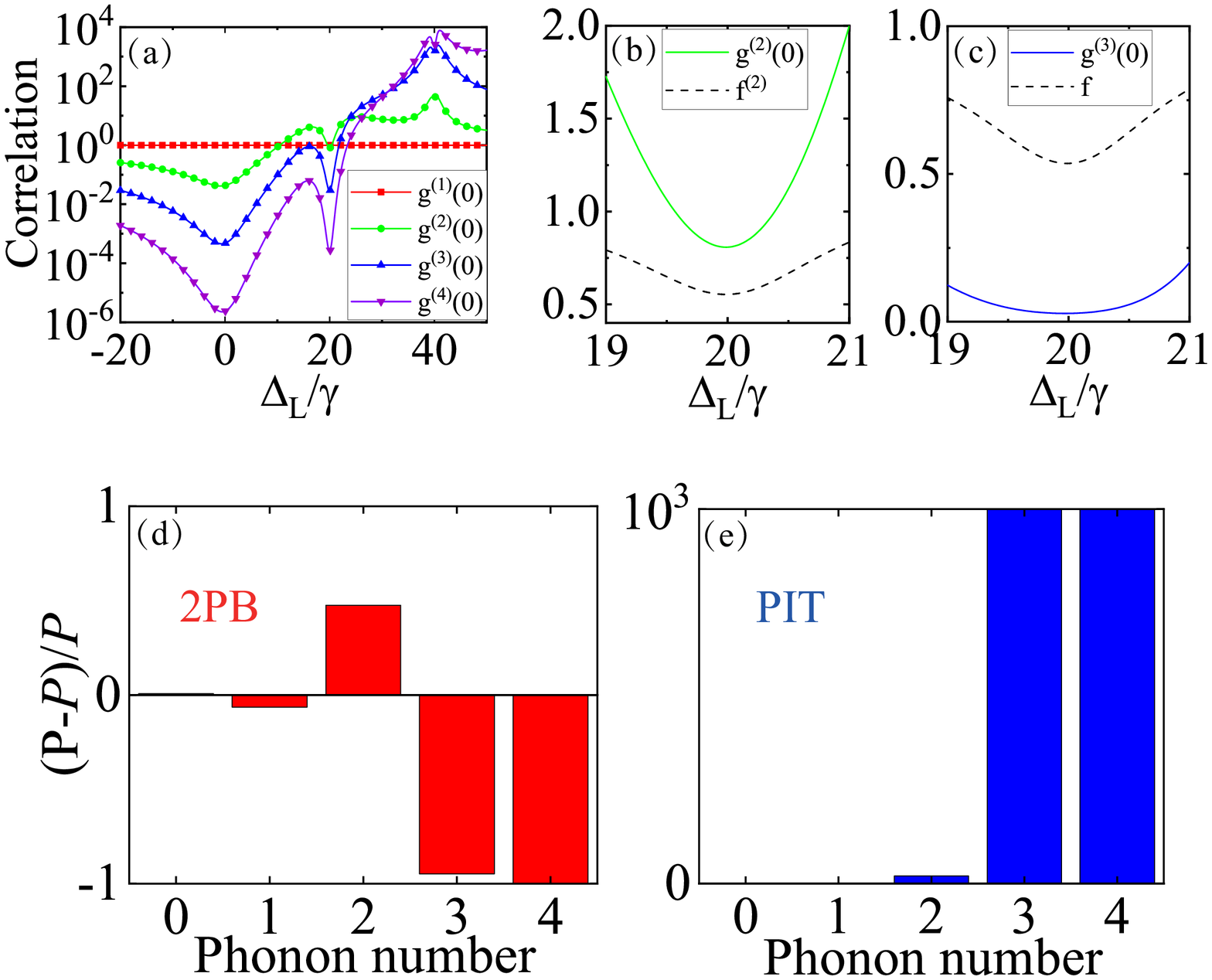}
\caption{\label{fig:4}(Color online)
(a) Under the strong driving region $\xi=3\gamma $, the correlation function versus the detuning $\Delta_{L}$ at $U=20\gamma$. (b) and (c) show the comparison for 2PB between the correlation function and the standard value in Eq. (\ref{ME19}). (d) and (e) show the deviation between the phonon population and the Poisson distributionfor 2PB amd PIT, respectively at $\Delta_{L}=20\gamma$ and $\Delta_{L}=40\gamma$.}
\end{figure}

Now we investigate PB under strong drive conditions depicted in Fig.~\ref{fig:4}. Figure 4(a), shows the results for the n-order correlation function versus the detuning $(\Delta_L)$ for $U=20\gamma$ and $\xi=3\gamma$. One can see the system still shows 1PB at $\Delta_{L}=0$. Unlike the aforementioned case in
Fig. 3(a), one may notice that there is second minimum in the correlation curves in Fig. 4(a), corresponding to 2 phonon blockade. The results of 2PB and PIT can be seen at $\Delta_{L}=20\gamma$ and $\Delta_{L}=40\gamma$, respectively.

According to Eq. (\ref{ME13}), 2PB should meet the following conditions for $n=2$:
\begin{eqnarray}\label{ME19}
	&(&i) \quad g^{(3)}(0)<\exp (-\langle\hat{m}\rangle) = f,\notag\\
	&(&ii) \quad g^{(2)}(0) \geq \exp (-\langle\hat{m}\rangle)+\langle\hat{m}\rangle \cdot g^{(3)}(0) = f^{(2)}.
\end{eqnarray}
In Fig.~\ref{fig:4}(b) and \ref{fig:4}(c) we show the correlation function versus the detuning $\Delta$, which satisfies the conditions in Eq. (\ref{ME19}). Further, in Fig.~\ref{fig:4}(d), we can see $P(2)>\mathcal{P}(2)$, $P(3)<\mathcal{P}(3)$, and $P(4)<\mathcal{P}(4)$, which meets our criterion given in Eq.~(\ref{ME16}) for $n=2$. Here, we can see that $P(3)>\mathcal{P}(3)$ and $P(4)>\mathcal{P}(4)$ in Fig.~\ref{fig:4}(e). Different from Fig. 3(a), here $g^{(4)}(0)>g^{(3)}(0)>g^{(2)}(0)>1$ in Fig.~\ref{fig:4}(a), corresponding to 3 phonon resonance, because of the significant increase in the drive strength.
It is worth mentioning that the PIT at $\xi=0.33\gamma$ (Fig. 3(a)) corresponds to the 2PB at $\xi=3\gamma$, which can explain the emergence of the hollow window of the maximum area in Fig.~\ref{fig:2}(b) and \ref{fig:2}(c), i.e., the transformation of PIT to PB by enhancing the driving strength. This can illustrate the equivalence of PB and PIT.

\section{Nonreciprocal effect}

\begin{figure}[b]
	\includegraphics[width=8.6cm]{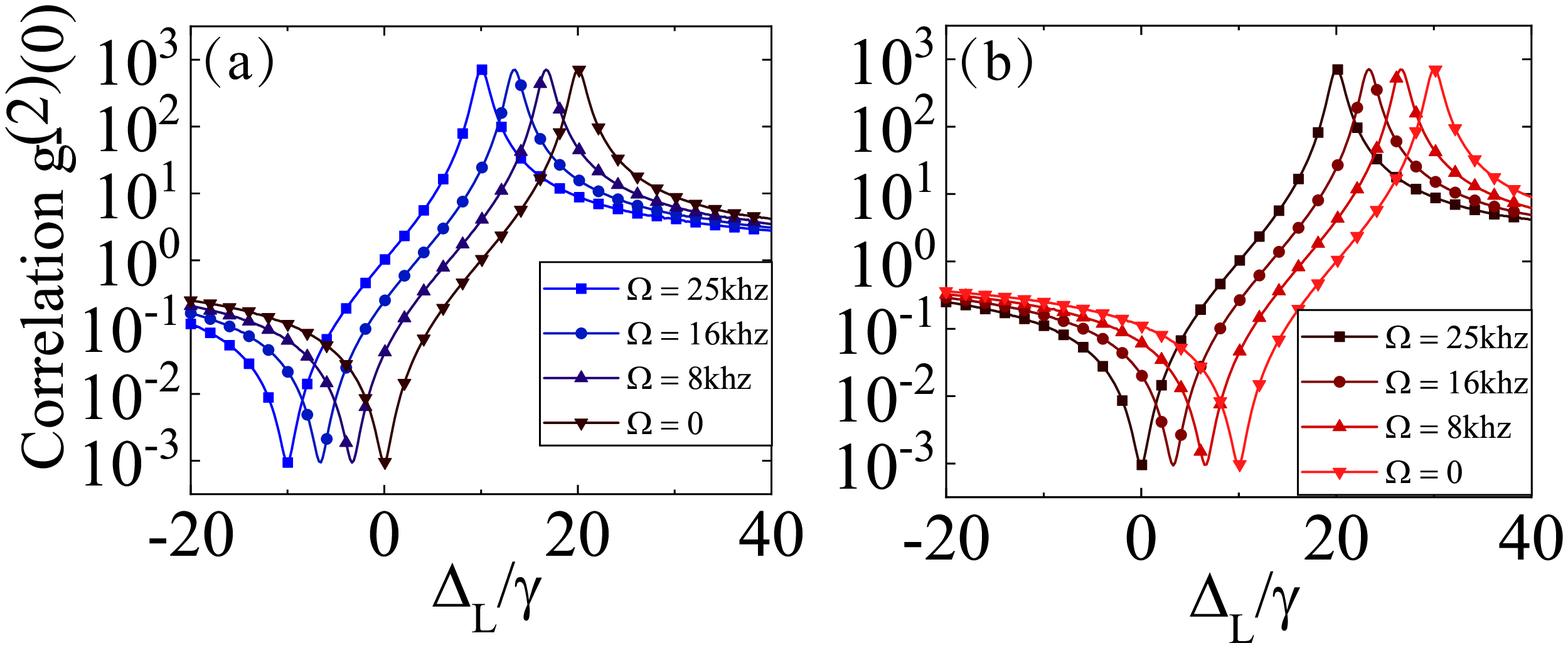}
	\caption{\label{fig:5}(Color online)
		The correlation function $g^{(2)}(0)$ versus the detuning $\Delta_{L}$ under weak driving $\xi=0.33\gamma$ and nonlinear constant $U=20\gamma$. (a) and (b) is corresponding to the results of driving the cavity from the right side and the left side of the waveguide, respectively.}
\end{figure}

Now, let's consider the effect of rotating the cavity. In Fig.~\ref{fig:5} we plot the correlation function $g^{(2)}(0)$ as a function of the detuning $\Delta_{L}$ when the angular velocity $\Omega$ takes various values in the weak driving regime.  In Fig.~\ref{fig:5}(a) (Fig.~\ref{fig:5}(b)) we can see the correlation curves shift to the left (right) after the angular velocity $\Omega$ increases.
\begin{figure}[t]
	\includegraphics[width=7cm, height=7cm]{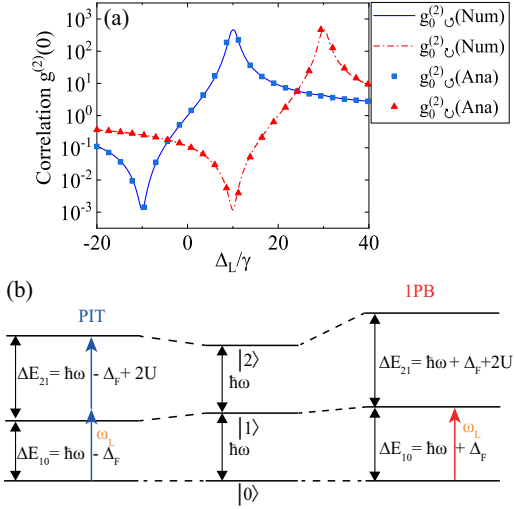}
	\caption{\label{fig:6}(Color online)
		(a) The correlation function $g^{(2)}(0)$ versus the detuning $\Delta_{L}$ with $\xi=0.33\gamma$, $\Delta_{F}=10\gamma$, and $U=20\gamma$. The chain (solid) line is the numerical solution result of driving the system from the left- (right-) side, and the triangle (square) scatter is the analytical solution result of driving the system from the left- (right-) side. (b) The energy levels of driving the system from the left-side and driving the system from the right-side become non-equidistant at $\Delta_{L}=10\gamma$.}
\end{figure}
Thus, the rotation of the cavity is the source of reciprocity, which causes the input driving from two different directions to have different resonance frequencies in the cavity.

By choosing the appropriate rotational angular velocity $\Omega$,  we can make different quantum effects appear at the same time. As shown in Fig.~\ref{fig:6}(a),  we have 1PB that appears when driving from one side, and PIT, which indicates that the absorption of the first phonon enhances the absorption of subsequent phonons appears when driving from the other side at $\Delta_{F}=\Delta_{L}=U/2=10\gamma$. Furthermore, Fig. \ref{fig:6}(a) shows that our analytical solution and the numerical solution are consistent; here we use $g_{\circlearrowright}^{(2)}(0)$  and $g_{\circlearrowleft}^{(2)}(0)$ to indicate that $\Delta_{F}>0$  and $\Delta_{F}<0$, respectively. We can clearly see that at $\Delta_{L}=10\gamma$, there is 1PB when driving the system from the left side, and PIT when driving the system from the right side, which is conformance to our expectation. For the difference of $g^{(2)}(0)$ for opposite directions, there is nonreciprocity with up to 6 orders of magnitude. This magnitude depends on the intensity of nonlinearity, which can be proved in Fig.~\ref{fig:2}. In Fig.~\ref{fig:6}(b), we can intuitively understand the reason why we have 1PB and PIT by considering the change of energy-level structure of the system.

\begin{figure}[b]
\includegraphics[width=7cm, height=6cm]{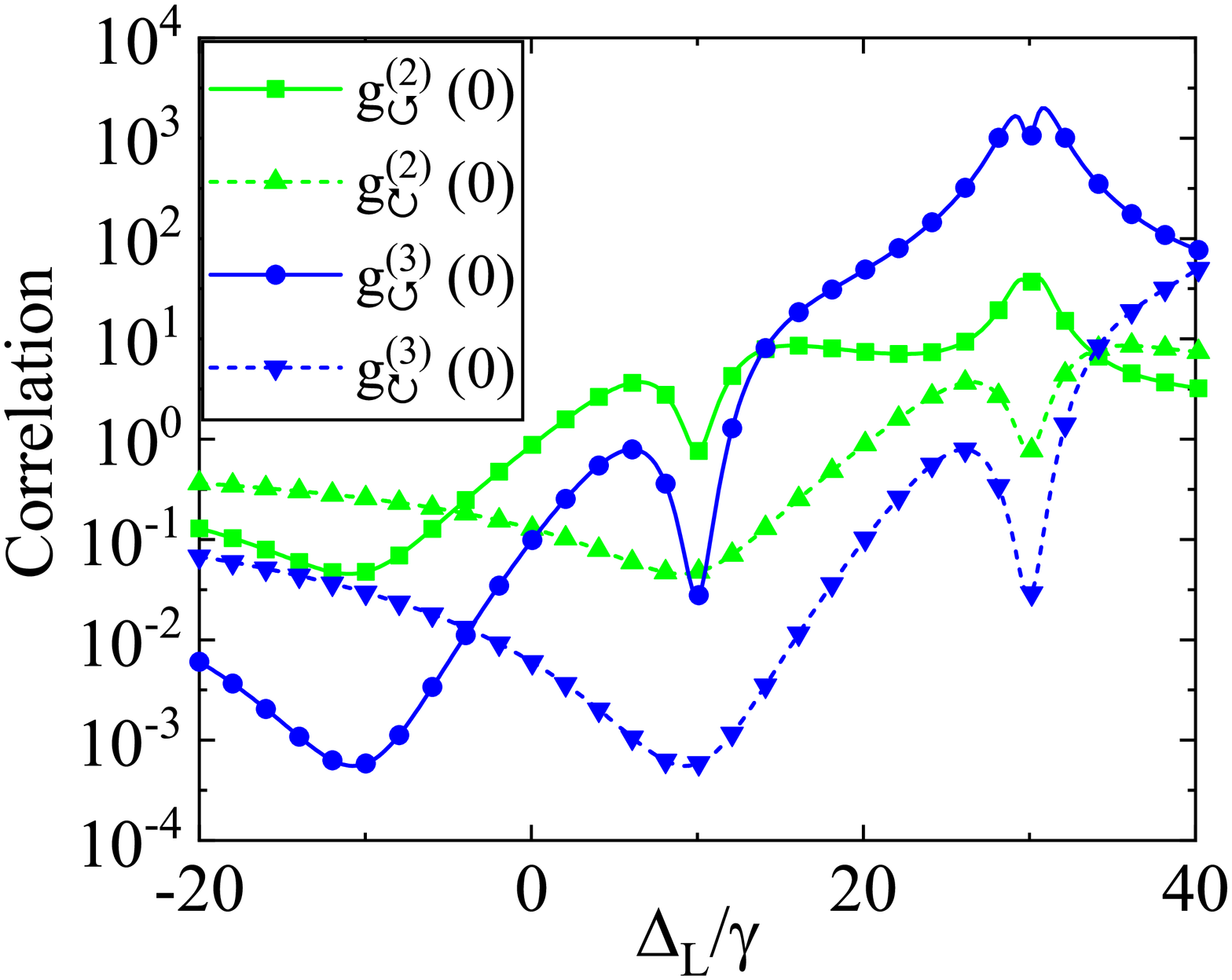}
\caption{\label{fig:7}(Color online)
The correlation function versus the detuning $\Delta_{L}$ with $\xi=3\gamma$, $\Delta_{F}=10\gamma$, and $U=20\gamma$. }
\end{figure}

\begin{figure}[t]
	\includegraphics[width=8cm]{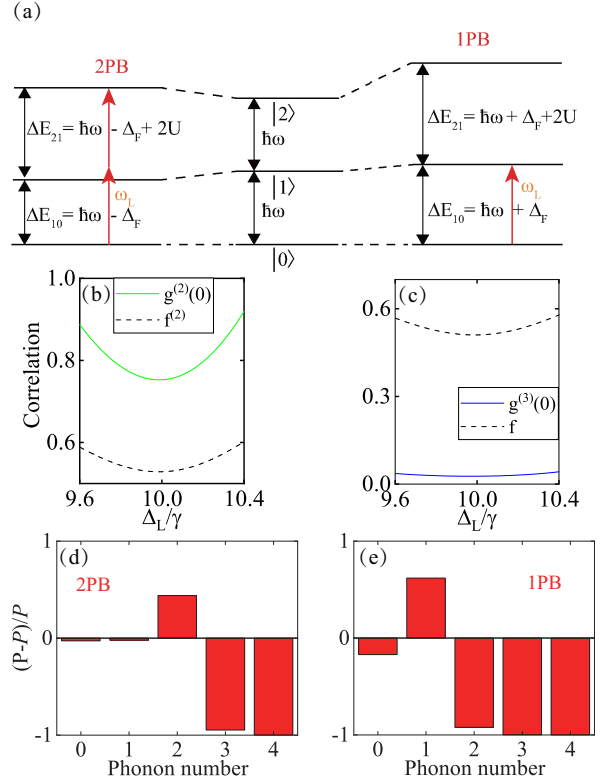}
	\caption{\label{fig:8}(Color online)
		(a) The energy levels of driving the system from the left- and driving the system from the right-side become non-equidistant for $\Delta_{L}=10\gamma$. (b) and (c) show the comparison for 2PB between the correlation function and the standard value in (\ref{ME19}), (d) and (e) show the deviation of the phonon population and the Poisson distribution for 1PB and 2PB, respectively.}
\end{figure}

\begin{figure}[t]
\includegraphics[width=8cm]{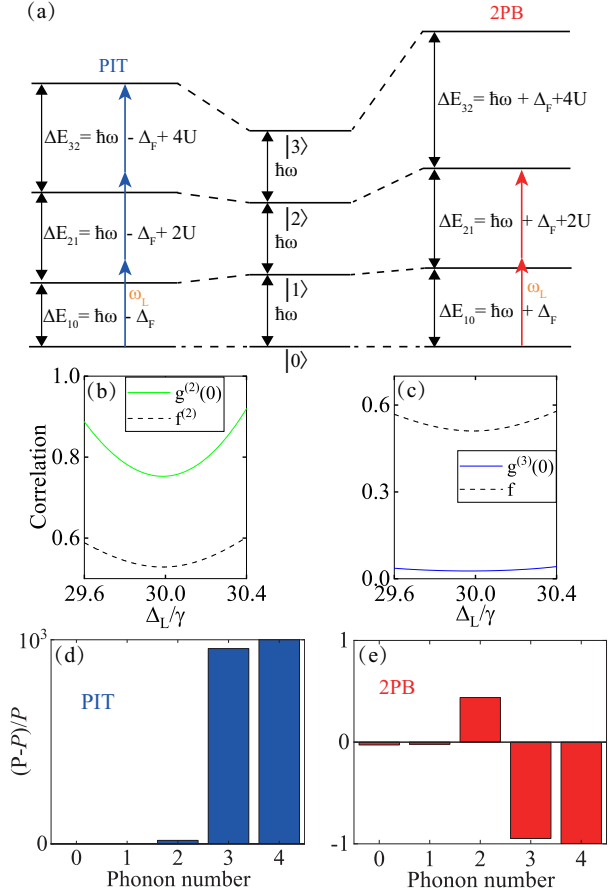}
\caption{\label{fig:9}(Color online)
(a) Under strong driving strength region $\xi=3 \gamma $, The energy levels of driving the system from the left- and driving the system from the right-side become non-equidistant for $\Delta_{L}=30\gamma$. (b) and (c) show the comparison for 2PB between the correlation function and the standard value in (\ref{ME19}), (d) and (e) show the deviation of the phonon population and the Poisson distribution for 2PB and PIT, respectively.}
\end{figure}

In Fig.~\ref{fig:7}, We show the correlation function versus the detuning $\Delta_{L}$ in the strong driving regime $(\xi=3\gamma)$ for $\Delta_{F}=10\gamma$. The solid curves represent the result of $g_{\circlearrowleft}^{(n)}(0)$, and the dashed line represents the result of $g_{\circlearrowright}^{(n)}(0)$. Here, at $\Delta_{L}=10\gamma$, we have 1PB and 2PB for the different driving directions. Furthermore, at $\Delta_{L}=30\gamma$, we obtain 2PB and PIT caused by the three-phonon resonance  for different driving directions. These results are discussed in detail in Fig.~\ref{fig:8} and Fig.~\ref{fig:9}.

In Fig.~\ref{fig:8}(a), we can intuitively understand the reason why we have 1PB and 2PB by considering the energy-level structure of the system. The plots in Fig.~\ref{fig:8}(b) and~\ref{fig:8}(c) show the comparison of correlation functions $(g^{2}(0)~\text{and}~g^{3}(0))$ with the standard values given in Eq.\ref{ME19} for strong driving strengths. We find that the curves of correlation functions satisfy the 2PB equation given in Eq. (\ref{ME19}). Moreover, in Fig.~\ref{fig:8}(d), we can find that $P(2)>\mathcal{P}(2)$, $P(3)<\mathcal{P}(3)$ and $P(4)<\mathcal{P}(4)$, which meets the criterion given in Eq. (\ref{ME16}) for $n=2$. In Fig.~\ref{fig:8}(e), we find that $P(1)>\mathcal{P}(1)$, $P(2)<\mathcal{P}(2)$ and $P(3)<\mathcal{P}(3)$, which meets the criterion in Eq.\ref{ME16} for $n=1$.

Next in Fig.~\ref{fig:9}(a), we show the energy-level structure of the system which enables us to understand the nonreciprocity of 2PB and PIT. Similarly in Fig.~\ref{fig:9}(b) and (c), we show that the curves of the correlation function satisfy the results given equation \ref{ME19}. Furthermore, we can see that $P(3)>\mathcal{P}(3)$ and $P(4)>\mathcal{P}(4)$ in Fig.~\ref{fig:9}(d). In Fig.~\ref{fig:9}(e), one can clearly see that $P(2)>\mathcal{P}(2)$, $P(3)<\mathcal{P}(3)$ and $P(4)<\mathcal{P}(4)$, which meets the criterion given in Eq. (\ref{ME16}) for $n=2$. In addition to the above results, we can also obtain nonreciprocity of 1PB and 3PR at $\Delta_{F}=20\gamma$ and $\Delta_{L}=20\gamma$,  nonreciprocity of 2PB and 4PR at  $\Delta_{F}=20\gamma$ and $\Delta_{L}=40\gamma$  for $U=20\gamma$.

\section{Experimental feasibility}
For examining the feasibility of this scheme in the experiment, we now discuss the relevant parameters. The acoustic ring cavity can be mounted on a rotating stage in the experiment \cite{96}. We can realize the coupling between the phononic waveguide and ring cavity by using wrap-around couplers. The dimensions of the cavity are $(r, w, h)=(50\ \mu \mathrm{m}, 5\ \mu \mathrm{m}, 0.5\ \mu \mathrm{m})$, and the mode volume is $V_{eff}=180\ \mu\mathrm{m}^3$\cite{113}. In our calculations, the resonance frequency of the cavity is $\omega/2\pi=3\ \mathrm{GHz}$, and the quality factor is $Q=10^7$. The frequency of phonons, which is in the $1\sim10\ \mathrm{GHz}$ range, can be easily implemented with the standard fabrication process.

For the SiV centers, the splitting of the sublevels of the lower orbital branch in the ground-state can be tuned by a few GHz by simply controlling local strain \cite{95}. The production scheme of the diamond membrane doped with SiV color centers is quite mature. The diamond membrane can not only be coupled with the cavity in contact on one side, and can also be placed above the cavity for coupling together~\cite{105}. If placed above the cavity, the diamond membrane can also be moved to change the coupling strength between the cavity mode and SiV center. The coupling strength of SiV centers and cavity is generally on the order of MHz{}. Thus the nonlinearity strength $U$ of $\mathrm{kHz}$ is easy to obtain. In our calculations, $U$ is in the $0\sim12\ \mathrm{kHz}$ range.

\section{CONCLUSION}
In conclusion, we study the nonreciprocal phonon blockade in a spinning acoustic ring resonator coupled to SiV centers in diamond. We show that in this hybrid  system, the SOI of phonons causes a nonreciprocal response. The statistical properties of phonons are exploited through calculating the higher-order correlation functions. We investigate the PB for various system parameters like the nonlinearity strength, driving strength, and drive frequency in a cavity rotating in the CCW direction. We find that in the weak driving regime,  nonreciprocity of 1PB and PIT for CW and CCW modes occurs, respectively. For strong driving we obtain nonreciprocity of 1PB and 2PB with the same system parameters. Furthermore, in the strong driving regime we have nonreciprocity of 2PB and PIT for CW and CCW modes, respectively. These results combined
with the current quantum technologies can be used to implement nonreciprocal few-phonon sources, nonreciprocal phonon routers and other quantum unidirectional devices.

\section*{Acknowledgments}
This work is
supported by the National Natural Science Foundation of
China under Grant No. 92065105  and
Natural Science Basic Research Program of Shaanxi
(Program No. 2020JC-02).


\begin{thebibliography}{107}%
\makeatletter
\providecommand \@ifxundefined [1]{%
 \@ifx{#1\undefined}
}%
\providecommand \@ifnum [1]{%
 \ifnum #1\expandafter \@firstoftwo
 \else \expandafter \@secondoftwo
 \fi
}%
\providecommand \@ifx [1]{%
 \ifx #1\expandafter \@firstoftwo
 \else \expandafter \@secondoftwo
 \fi
}%
\providecommand \natexlab [1]{#1}%
\providecommand \enquote  [1]{``#1''}%
\providecommand \bibnamefont  [1]{#1}%
\providecommand \bibfnamefont [1]{#1}%
\providecommand \citenamefont [1]{#1}%
\providecommand \href@noop [0]{\@secondoftwo}%
\providecommand \href [0]{\begingroup \@sanitize@url \@href}%
\providecommand \@href[1]{\@@startlink{#1}\@@href}%
\providecommand \@@href[1]{\endgroup#1\@@endlink}%
\providecommand \@sanitize@url [0]{\catcode `\\12\catcode `\$12\catcode
  `\&12\catcode `\#12\catcode `\^12\catcode `\_12\catcode `\%12\relax}%
\providecommand \@@startlink[1]{}%
\providecommand \@@endlink[0]{}%
\providecommand \url  [0]{\begingroup\@sanitize@url \@url }%
\providecommand \@url [1]{\endgroup\@href {#1}{\urlprefix }}%
\providecommand \urlprefix  [0]{URL }%
\providecommand \Eprint [0]{\href }%
\providecommand \doibase [0]{https://doi.org/}%
\providecommand \selectlanguage [0]{\@gobble}%
\providecommand \bibinfo  [0]{\@secondoftwo}%
\providecommand \bibfield  [0]{\@secondoftwo}%
\providecommand \translation [1]{[#1]}%
\providecommand \BibitemOpen [0]{}%
\providecommand \bibitemStop [0]{}%
\providecommand \bibitemNoStop [0]{.\EOS\space}%
\providecommand \EOS [0]{\spacefactor3000\relax}%
\providecommand \BibitemShut  [1]{\csname bibitem#1\endcsname}%
\let\auto@bib@innerbib\@empty
\bibitem [{\citenamefont {Palomaki}\ \emph {et~al.}(2013)\citenamefont
  {Palomaki}, \citenamefont {Teufel}, \citenamefont {Simmonds},\ and\
  \citenamefont {Lehnert}}]{1}%
  \BibitemOpen
  \bibfield  {author} {\bibinfo {author} {\bibfnamefont {T.~A.}\ \bibnamefont
  {Palomaki}}, \bibinfo {author} {\bibfnamefont {J.~D.}\ \bibnamefont
  {Teufel}}, \bibinfo {author} {\bibfnamefont {R.~W.}\ \bibnamefont
  {Simmonds}},\ and\ \bibinfo {author} {\bibfnamefont {K.~W.}\ \bibnamefont
  {Lehnert}},\ }\bibfield  {title} {\bibinfo {title} {Entangling mechanical
  motion with microwave fields},\ }\href
  {https://doi.org/10.1126/science.1244563} {\bibfield  {journal} {\bibinfo
  {journal} {Science}\ }\textbf {\bibinfo {volume} {342}},\ \bibinfo {pages}
  {710} (\bibinfo {year} {2013})}\BibitemShut {NoStop}%
\bibitem [{\citenamefont {Okamoto}\ \emph {et~al.}(2013)\citenamefont
  {Okamoto}, \citenamefont {Gourgout}, \citenamefont {Chang}, \citenamefont
  {Onomitsu}, \citenamefont {Mahboob}, \citenamefont {Chang},\ and\
  \citenamefont {Yamaguchi}}]{2}%
  \BibitemOpen
  \bibfield  {author} {\bibinfo {author} {\bibfnamefont {H.}~\bibnamefont
  {Okamoto}}, \bibinfo {author} {\bibfnamefont {A.}~\bibnamefont {Gourgout}},
  \bibinfo {author} {\bibfnamefont {C.~Y.}\ \bibnamefont {Chang}}, \bibinfo
  {author} {\bibfnamefont {K.}~\bibnamefont {Onomitsu}}, \bibinfo {author}
  {\bibfnamefont {I.}~\bibnamefont {Mahboob}}, \bibinfo {author} {\bibfnamefont
  {E.~Y.}\ \bibnamefont {Chang}},\ and\ \bibinfo {author} {\bibfnamefont
  {H.}~\bibnamefont {Yamaguchi}},\ }\bibfield  {title} {\bibinfo {title}
  {Coherent phonon manipulation in coupled mechanical resonators},\ }\href
  {https://doi.org/10.1038/Nphys2665} {\bibfield  {journal} {\bibinfo
  {journal} {Nat. Phys.}\ }\textbf {\bibinfo {volume} {9}},\ \bibinfo {pages}
  {480} (\bibinfo {year} {2013})}\BibitemShut {NoStop}%
\bibitem [{\citenamefont {Jing}\ \emph {et~al.}(2014)\citenamefont {Jing},
  \citenamefont {Ozdemir}, \citenamefont {Lu}, \citenamefont {Zhang},
  \citenamefont {Yang},\ and\ \citenamefont {Nori}}]{3}%
  \BibitemOpen
  \bibfield  {author} {\bibinfo {author} {\bibfnamefont {H.}~\bibnamefont
  {Jing}}, \bibinfo {author} {\bibfnamefont {S.~K.}\ \bibnamefont {Ozdemir}},
  \bibinfo {author} {\bibfnamefont {X.~Y.}\ \bibnamefont {Lu}}, \bibinfo
  {author} {\bibfnamefont {J.}~\bibnamefont {Zhang}}, \bibinfo {author}
  {\bibfnamefont {L.}~\bibnamefont {Yang}},\ and\ \bibinfo {author}
  {\bibfnamefont {F.}~\bibnamefont {Nori}},\ }\bibfield  {title} {\bibinfo
  {title} {Pt-symmetric phonon laser},\ }\href
  {https://doi.org/10.1103/PhysRevLett.113.053604} {\bibfield  {journal}
  {\bibinfo  {journal} {Phys. Rev. Lett.}\ }\textbf {\bibinfo {volume} {113}},\
  \bibinfo {pages} {053604} (\bibinfo {year} {2014})}\BibitemShut {NoStop}%
\bibitem [{\citenamefont {Hong}\ \emph {et~al.}(2017)\citenamefont {Hong},
  \citenamefont {Riedinger}, \citenamefont {Marinkovic}, \citenamefont
  {Wallucks}, \citenamefont {Hofer}, \citenamefont {Norte}, \citenamefont
  {Aspelmeyer},\ and\ \citenamefont {Groblacher}}]{4}%
  \BibitemOpen
  \bibfield  {author} {\bibinfo {author} {\bibfnamefont {S.}~\bibnamefont
  {Hong}}, \bibinfo {author} {\bibfnamefont {R.}~\bibnamefont {Riedinger}},
  \bibinfo {author} {\bibfnamefont {I.}~\bibnamefont {Marinkovic}}, \bibinfo
  {author} {\bibfnamefont {A.}~\bibnamefont {Wallucks}}, \bibinfo {author}
  {\bibfnamefont {S.~G.}\ \bibnamefont {Hofer}}, \bibinfo {author}
  {\bibfnamefont {R.~A.}\ \bibnamefont {Norte}}, \bibinfo {author}
  {\bibfnamefont {M.}~\bibnamefont {Aspelmeyer}},\ and\ \bibinfo {author}
  {\bibfnamefont {S.}~\bibnamefont {Groblacher}},\ }\bibfield  {title}
  {\bibinfo {title} {Hanbury brown and twiss interferometry of single phonons
  from an optomechanical resonator},\ }\href
  {https://doi.org/10.1126/science.aan7939} {\bibfield  {journal} {\bibinfo
  {journal} {Science}\ }\textbf {\bibinfo {volume} {358}},\ \bibinfo {pages}
  {203} (\bibinfo {year} {2017})}\BibitemShut {NoStop}%
\bibitem [{\citenamefont {Luo}\ \emph {et~al.}(2018)\citenamefont {Luo},
  \citenamefont {Zhang}, \citenamefont {Deng}, \citenamefont {Li},
  \citenamefont {Cao}, \citenamefont {Xiao}, \citenamefont {Guo}, \citenamefont
  {Tian},\ and\ \citenamefont {Guo}}]{5}%
  \BibitemOpen
  \bibfield  {author} {\bibinfo {author} {\bibfnamefont {G.}~\bibnamefont
  {Luo}}, \bibinfo {author} {\bibfnamefont {Z.-Z.}\ \bibnamefont {Zhang}},
  \bibinfo {author} {\bibfnamefont {G.-W.}\ \bibnamefont {Deng}}, \bibinfo
  {author} {\bibfnamefont {H.-O.}\ \bibnamefont {Li}}, \bibinfo {author}
  {\bibfnamefont {G.}~\bibnamefont {Cao}}, \bibinfo {author} {\bibfnamefont
  {M.}~\bibnamefont {Xiao}}, \bibinfo {author} {\bibfnamefont {G.-C.}\
  \bibnamefont {Guo}}, \bibinfo {author} {\bibfnamefont {L.}~\bibnamefont
  {Tian}},\ and\ \bibinfo {author} {\bibfnamefont {G.-P.}\ \bibnamefont
  {Guo}},\ }\bibfield  {title} {\bibinfo {title} {Strong indirect coupling
  between graphene-based mechanical resonators via a phonon cavity},\ }\href
  {https://doi.org/10.1038/s41467-018-02854-4} {\bibfield  {journal} {\bibinfo
  {journal} {Nat. Commun.}\ }\textbf {\bibinfo {volume} {9}},\ \bibinfo {pages}
  {383} (\bibinfo {year} {2018})}\BibitemShut {NoStop}%
\bibitem [{\citenamefont {Merklein}\ \emph {et~al.}(2015)\citenamefont
  {Merklein}, \citenamefont {Kabakova}, \citenamefont {Büttner}, \citenamefont
  {Choi}, \citenamefont {Luther-Davies}, \citenamefont {Madden},\ and\
  \citenamefont {Eggleton}}]{6}%
  \BibitemOpen
  \bibfield  {author} {\bibinfo {author} {\bibfnamefont {M.}~\bibnamefont
  {Merklein}}, \bibinfo {author} {\bibfnamefont {I.~V.}\ \bibnamefont
  {Kabakova}}, \bibinfo {author} {\bibfnamefont {T.~F.~S.}\ \bibnamefont
  {Büttner}}, \bibinfo {author} {\bibfnamefont {D.-Y.}\ \bibnamefont {Choi}},
  \bibinfo {author} {\bibfnamefont {B.}~\bibnamefont {Luther-Davies}}, \bibinfo
  {author} {\bibfnamefont {S.~J.}\ \bibnamefont {Madden}},\ and\ \bibinfo
  {author} {\bibfnamefont {B.~J.}\ \bibnamefont {Eggleton}},\ }\bibfield
  {title} {\bibinfo {title} {Enhancing and inhibiting stimulated brillouin
  scattering in photonic integrated circuits},\ }\href
  {https://doi.org/10.1038/ncomms7396} {\bibfield  {journal} {\bibinfo
  {journal} {Nat. Commun.}\ }\textbf {\bibinfo {volume} {6}},\ \bibinfo {pages}
  {6396} (\bibinfo {year} {2015})}\BibitemShut {NoStop}%
\bibitem [{\citenamefont {Van~Laer}\ \emph {et~al.}(2015)\citenamefont
  {Van~Laer}, \citenamefont {Kuyken}, \citenamefont {Van~Thourhout},\ and\
  \citenamefont {Baets}}]{7}%
  \BibitemOpen
  \bibfield  {author} {\bibinfo {author} {\bibfnamefont {R.}~\bibnamefont
  {Van~Laer}}, \bibinfo {author} {\bibfnamefont {B.}~\bibnamefont {Kuyken}},
  \bibinfo {author} {\bibfnamefont {D.}~\bibnamefont {Van~Thourhout}},\ and\
  \bibinfo {author} {\bibfnamefont {R.}~\bibnamefont {Baets}},\ }\bibfield
  {title} {\bibinfo {title} {Interaction between light and highly confined
  hypersound in a silicon photonic nanowire},\ }\href
  {https://doi.org/10.1038/Nphoton.2015.11} {\bibfield  {journal} {\bibinfo
  {journal} {Nat. Photon.}\ }\textbf {\bibinfo {volume} {9}},\ \bibinfo {pages}
  {199} (\bibinfo {year} {2015})}\BibitemShut {NoStop}%
\bibitem [{\citenamefont {Li}\ \emph {et~al.}(2016)\citenamefont {Li},
  \citenamefont {Xiang}, \citenamefont {Rabl},\ and\ \citenamefont {Nori}}]{8}%
  \BibitemOpen
  \bibfield  {author} {\bibinfo {author} {\bibfnamefont {P.-B.}\ \bibnamefont
  {Li}}, \bibinfo {author} {\bibfnamefont {Z.-L.}\ \bibnamefont {Xiang}},
  \bibinfo {author} {\bibfnamefont {P.}~\bibnamefont {Rabl}},\ and\ \bibinfo
  {author} {\bibfnamefont {F.}~\bibnamefont {Nori}},\ }\bibfield  {title}
  {\bibinfo {title} {Hybrid quantum device with nitrogen-vacancy centers in
  diamond coupled to carbon nanotubes},\ }\href
  {https://doi.org/10.1103/PhysRevLett.117.015502} {\bibfield  {journal}
  {\bibinfo  {journal} {Phys. Rev. Lett.}\ }\textbf {\bibinfo {volume} {117}},\
  \bibinfo {pages} {015502} (\bibinfo {year} {2016})}\BibitemShut {NoStop}%
\bibitem [{\citenamefont {Kuzyk}\ and\ \citenamefont {Wang}(2018)}]{9}%
  \BibitemOpen
  \bibfield  {author} {\bibinfo {author} {\bibfnamefont {M.~C.}\ \bibnamefont
  {Kuzyk}}\ and\ \bibinfo {author} {\bibfnamefont {H.}~\bibnamefont {Wang}},\
  }\bibfield  {title} {\bibinfo {title} {Scaling phononic quantum networks of
  solid-state spins with closed mechanical subsystems},\ }\href
  {https://doi.org/10.1103/PhysRevX.8.041027} {\bibfield  {journal} {\bibinfo
  {journal} {Phys. Rev. X}\ }\textbf {\bibinfo {volume} {8}},\ \bibinfo {pages}
  {041027} (\bibinfo {year} {2018})}\BibitemShut {NoStop}%
\bibitem [{\citenamefont {Ghadimi}\ \emph {et~al.}(2018)\citenamefont
  {Ghadimi}, \citenamefont {Fedorov}, \citenamefont {Engelsen}, \citenamefont
  {Bereyhi}, \citenamefont {Schilling}, \citenamefont {Wilson},\ and\
  \citenamefont {Kippenberg}}]{10}%
  \BibitemOpen
  \bibfield  {author} {\bibinfo {author} {\bibfnamefont {A.~H.}\ \bibnamefont
  {Ghadimi}}, \bibinfo {author} {\bibfnamefont {S.~A.}\ \bibnamefont
  {Fedorov}}, \bibinfo {author} {\bibfnamefont {N.~J.}\ \bibnamefont
  {Engelsen}}, \bibinfo {author} {\bibfnamefont {M.~J.}\ \bibnamefont
  {Bereyhi}}, \bibinfo {author} {\bibfnamefont {R.}~\bibnamefont {Schilling}},
  \bibinfo {author} {\bibfnamefont {D.~J.}\ \bibnamefont {Wilson}},\ and\
  \bibinfo {author} {\bibfnamefont {T.~J.}\ \bibnamefont {Kippenberg}},\
  }\bibfield  {title} {\bibinfo {title} {Elastic strain engineering for
  ultralow mechanical dissipation},\ }\href
  {https://doi.org/10.1126/science.aar6939} {\bibfield  {journal} {\bibinfo
  {journal} {Science}\ }\textbf {\bibinfo {volume} {360}},\ \bibinfo {pages}
  {764} (\bibinfo {year} {2018})}\BibitemShut {NoStop}%
\bibitem [{\citenamefont {Bin}\ \emph {et~al.}(2020)\citenamefont {Bin},
  \citenamefont {Lü}, \citenamefont {Laussy}, \citenamefont {Nori},\ and\
  \citenamefont {Wu}}]{11}%
  \BibitemOpen
  \bibfield  {author} {\bibinfo {author} {\bibfnamefont {Q.}~\bibnamefont
  {Bin}}, \bibinfo {author} {\bibfnamefont {X.-Y.}\ \bibnamefont {Lü}},
  \bibinfo {author} {\bibfnamefont {F.~P.}\ \bibnamefont {Laussy}}, \bibinfo
  {author} {\bibfnamefont {F.}~\bibnamefont {Nori}},\ and\ \bibinfo {author}
  {\bibfnamefont {Y.}~\bibnamefont {Wu}},\ }\bibfield  {title} {\bibinfo
  {title} {$n$-phonon bundle emission via the stokes process},\ }\href
  {https://doi.org/10.1103/PhysRevLett.124.053601} {\bibfield  {journal}
  {\bibinfo  {journal} {Phys. Rev. Lett.}\ }\textbf {\bibinfo {volume} {124}},\
  \bibinfo {pages} {053601} (\bibinfo {year} {2020})}\BibitemShut {NoStop}%
\bibitem [{\citenamefont {Dong}\ and\ \citenamefont {Li}(2019)}]{12}%
  \BibitemOpen
  \bibfield  {author} {\bibinfo {author} {\bibfnamefont {X.-L.}\ \bibnamefont
  {Dong}}\ and\ \bibinfo {author} {\bibfnamefont {P.-B.}\ \bibnamefont {Li}},\
  }\bibfield  {title} {\bibinfo {title} {Multiphonon interactions between
  nitrogen-vacancy centers and nanomechanical resonators},\ }\href
  {https://doi.org/10.1103/PhysRevA.100.043825} {\bibfield  {journal} {\bibinfo
   {journal} {Phys. Rev. A}\ }\textbf {\bibinfo {volume} {100}},\ \bibinfo
  {pages} {043825} (\bibinfo {year} {2019})}\BibitemShut {NoStop}%
\bibitem [{\citenamefont {Liu}\ \emph {et~al.}(2010)\citenamefont {Liu},
  \citenamefont {Miranowicz}, \citenamefont {Gao}, \citenamefont {Bajer},
  \citenamefont {Sun},\ and\ \citenamefont {Nori}}]{13}%
  \BibitemOpen
  \bibfield  {author} {\bibinfo {author} {\bibfnamefont {Y.-x.}\ \bibnamefont
  {Liu}}, \bibinfo {author} {\bibfnamefont {A.}~\bibnamefont {Miranowicz}},
  \bibinfo {author} {\bibfnamefont {Y.~B.}\ \bibnamefont {Gao}}, \bibinfo
  {author} {\bibfnamefont {J.}~\bibnamefont {Bajer}}, \bibinfo {author}
  {\bibfnamefont {C.~P.}\ \bibnamefont {Sun}},\ and\ \bibinfo {author}
  {\bibfnamefont {F.}~\bibnamefont {Nori}},\ }\bibfield  {title} {\bibinfo
  {title} {Qubit-induced phonon blockade as a signature of quantum behavior in
  nanomechanical resonators},\ }\href
  {https://doi.org/10.1103/PhysRevA.82.032101} {\bibfield  {journal} {\bibinfo
  {journal} {Phys. Rev. A}\ }\textbf {\bibinfo {volume} {82}},\ \bibinfo
  {pages} {032101} (\bibinfo {year} {2010})}\BibitemShut {NoStop}%
\bibitem [{\citenamefont {Kastner}(1993)}]{14}%
  \BibitemOpen
  \bibfield  {author} {\bibinfo {author} {\bibfnamefont {M.~A.}\ \bibnamefont
  {Kastner}},\ }\bibfield  {title} {\bibinfo {title} {Artificial atoms},\
  }\href {https://doi.org/10.1063/1.881393} {\bibfield  {journal} {\bibinfo
  {journal} {Phys. Today}\ }\textbf {\bibinfo {volume} {46}},\ \bibinfo {pages}
  {24} (\bibinfo {year} {1993})}\BibitemShut {NoStop}%
\bibitem [{\citenamefont {Leonski}\ and\ \citenamefont {Tanas}(1994)}]{15}%
  \BibitemOpen
  \bibfield  {author} {\bibinfo {author} {\bibfnamefont {W.}~\bibnamefont
  {Leonski}}\ and\ \bibinfo {author} {\bibfnamefont {R.}~\bibnamefont
  {Tanas}},\ }\bibfield  {title} {\bibinfo {title} {Possibility of producing
  the one-photon state in a kicked cavity with a nonlinear kerr medium},\
  }\href {https://doi.org/10.1103/PhysRevA.49.R20} {\bibfield  {journal}
  {\bibinfo  {journal} {Phys. Rev. A}\ }\textbf {\bibinfo {volume} {49}},\
  \bibinfo {pages} {R20} (\bibinfo {year} {1994})}\BibitemShut {NoStop}%
\bibitem [{\citenamefont {Imamoglu}\ \emph {et~al.}(1997)\citenamefont
  {Imamoglu}, \citenamefont {Schmidt}, \citenamefont {Woods},\ and\
  \citenamefont {Deutsch}}]{16}%
  \BibitemOpen
  \bibfield  {author} {\bibinfo {author} {\bibfnamefont {A.}~\bibnamefont
  {Imamoglu}}, \bibinfo {author} {\bibfnamefont {H.}~\bibnamefont {Schmidt}},
  \bibinfo {author} {\bibfnamefont {G.}~\bibnamefont {Woods}},\ and\ \bibinfo
  {author} {\bibfnamefont {M.}~\bibnamefont {Deutsch}},\ }\bibfield  {title}
  {\bibinfo {title} {Strongly interacting photons in a nonlinear cavity},\
  }\href {https://doi.org/10.1103/PhysRevLett.79.1467} {\bibfield  {journal}
  {\bibinfo  {journal} {Phys. Rev. Lett.}\ }\textbf {\bibinfo {volume} {79}},\
  \bibinfo {pages} {1467} (\bibinfo {year} {1997})}\BibitemShut {NoStop}%
\bibitem [{\citenamefont {Leonski}(1997)}]{17}%
  \BibitemOpen
  \bibfield  {author} {\bibinfo {author} {\bibfnamefont {W.}~\bibnamefont
  {Leonski}},\ }\bibfield  {title} {\bibinfo {title} {Finite-dimensional
  coherent-state generation and quantum-optical nonlinear oscillator models},\
  }\href {https://doi.org/10.1103/PhysRevA.55.3874} {\bibfield  {journal}
  {\bibinfo  {journal} {Phys. Rev. A}\ }\textbf {\bibinfo {volume} {55}},\
  \bibinfo {pages} {3874} (\bibinfo {year} {1997})}\BibitemShut {NoStop}%
\bibitem [{\citenamefont {Rabl}(2011)}]{18}%
  \BibitemOpen
  \bibfield  {author} {\bibinfo {author} {\bibfnamefont {P.}~\bibnamefont
  {Rabl}},\ }\bibfield  {title} {\bibinfo {title} {Photon blockade effect in
  optomechanical systems},\ }\href
  {https://doi.org/10.1103/PhysRevLett.107.063601} {\bibfield  {journal}
  {\bibinfo  {journal} {Phys. Rev. Lett.}\ }\textbf {\bibinfo {volume} {107}},\
  \bibinfo {pages} {063601} (\bibinfo {year} {2011})}\BibitemShut {NoStop}%
\bibitem [{\citenamefont {Xu}\ \emph {et~al.}(2013)\citenamefont {Xu},
  \citenamefont {Li},\ and\ \citenamefont {Liu}}]{19}%
  \BibitemOpen
  \bibfield  {author} {\bibinfo {author} {\bibfnamefont {X.-W.}\ \bibnamefont
  {Xu}}, \bibinfo {author} {\bibfnamefont {Y.-J.}\ \bibnamefont {Li}},\ and\
  \bibinfo {author} {\bibfnamefont {Y.-x.}\ \bibnamefont {Liu}},\ }\bibfield
  {title} {\bibinfo {title} {Photon-induced tunneling in optomechanical
  systems},\ }\href {https://doi.org/10.1103/PhysRevA.87.025803} {\bibfield
  {journal} {\bibinfo  {journal} {Phys. Rev. A}\ }\textbf {\bibinfo {volume}
  {87}},\ \bibinfo {pages} {025803} (\bibinfo {year} {2013})}\BibitemShut
  {NoStop}%
\bibitem [{\citenamefont {Wang}\ \emph {et~al.}(2019)\citenamefont {Wang},
  \citenamefont {Bai}, \citenamefont {Liu}, \citenamefont {Zhang},\ and\
  \citenamefont {Wang}}]{20}%
  \BibitemOpen
  \bibfield  {author} {\bibinfo {author} {\bibfnamefont {D.-Y.}\ \bibnamefont
  {Wang}}, \bibinfo {author} {\bibfnamefont {C.-H.}\ \bibnamefont {Bai}},
  \bibinfo {author} {\bibfnamefont {S.}~\bibnamefont {Liu}}, \bibinfo {author}
  {\bibfnamefont {S.}~\bibnamefont {Zhang}},\ and\ \bibinfo {author}
  {\bibfnamefont {H.-F.}\ \bibnamefont {Wang}},\ }\bibfield  {title} {\bibinfo
  {title} {Distinguishing photon blockade in a $\mathcal{PT}$-symmetric
  optomechanical system},\ }\href {https://doi.org/10.1103/PhysRevA.99.043818}
  {\bibfield  {journal} {\bibinfo  {journal} {Phys. Rev. A}\ }\textbf {\bibinfo
  {volume} {99}},\ \bibinfo {pages} {043818} (\bibinfo {year}
  {2019})}\BibitemShut {NoStop}%
\bibitem [{\citenamefont {Wang}\ \emph
  {et~al.}(2020{\natexlab{a}})\citenamefont {Wang}, \citenamefont {Gao},
  \citenamefont {Pang}, \citenamefont {Wang},\ and\ \citenamefont {Wang}}]{21}%
  \BibitemOpen
  \bibfield  {author} {\bibinfo {author} {\bibfnamefont {K.}~\bibnamefont
  {Wang}}, \bibinfo {author} {\bibfnamefont {Y.-P.}\ \bibnamefont {Gao}},
  \bibinfo {author} {\bibfnamefont {T.-T.}\ \bibnamefont {Pang}}, \bibinfo
  {author} {\bibfnamefont {T.-J.}\ \bibnamefont {Wang}},\ and\ \bibinfo
  {author} {\bibfnamefont {C.}~\bibnamefont {Wang}},\ }\bibfield  {title}
  {\bibinfo {title} {Enhanced photon blockade in quadratically coupled
  optomechanical system},\ }\href {https://doi.org/10.1209/0295-5075/131/24003}
  {\bibfield  {journal} {\bibinfo  {journal} {EPL}\ }\textbf {\bibinfo {volume}
  {131}},\ \bibinfo {pages} {24003} (\bibinfo {year}
  {2020}{\natexlab{a}})}\BibitemShut {NoStop}%
\bibitem [{\citenamefont {Wang}\ \emph
  {et~al.}(2020{\natexlab{b}})\citenamefont {Wang}, \citenamefont {Bai},
  \citenamefont {Liu}, \citenamefont {Zhang},\ and\ \citenamefont {Wang}}]{22}%
  \BibitemOpen
  \bibfield  {author} {\bibinfo {author} {\bibfnamefont {D.-Y.}\ \bibnamefont
  {Wang}}, \bibinfo {author} {\bibfnamefont {C.-H.}\ \bibnamefont {Bai}},
  \bibinfo {author} {\bibfnamefont {S.}~\bibnamefont {Liu}}, \bibinfo {author}
  {\bibfnamefont {S.}~\bibnamefont {Zhang}},\ and\ \bibinfo {author}
  {\bibfnamefont {H.-F.}\ \bibnamefont {Wang}},\ }\bibfield  {title} {\bibinfo
  {title} {Photon blockade in a double-cavity optomechanical system with
  nonreciprocal coupling},\ }\href {https://doi.org/10.1088/1367-2630/abaa8a}
  {\bibfield  {journal} {\bibinfo  {journal} {New J. Phys.}\ }\textbf {\bibinfo
  {volume} {22}},\ \bibinfo {pages} {093006} (\bibinfo {year}
  {2020}{\natexlab{b}})}\BibitemShut {NoStop}%
\bibitem [{\citenamefont {Didier}\ \emph {et~al.}(2011)\citenamefont {Didier},
  \citenamefont {Pugnetti}, \citenamefont {Blanter},\ and\ \citenamefont
  {Fazio}}]{23}%
  \BibitemOpen
  \bibfield  {author} {\bibinfo {author} {\bibfnamefont {N.}~\bibnamefont
  {Didier}}, \bibinfo {author} {\bibfnamefont {S.}~\bibnamefont {Pugnetti}},
  \bibinfo {author} {\bibfnamefont {Y.~M.}\ \bibnamefont {Blanter}},\ and\
  \bibinfo {author} {\bibfnamefont {R.}~\bibnamefont {Fazio}},\ }\bibfield
  {title} {\bibinfo {title} {Detecting phonon blockade with photons},\ }\href
  {https://doi.org/10.1103/PhysRevB.84.054503} {\bibfield  {journal} {\bibinfo
  {journal} {Phys. Rev. B}\ }\textbf {\bibinfo {volume} {84}},\ \bibinfo
  {pages} {054503} (\bibinfo {year} {2011})}\BibitemShut {NoStop}%
\bibitem [{\citenamefont {Ramos}\ \emph {et~al.}(2013)\citenamefont {Ramos},
  \citenamefont {Sudhir}, \citenamefont {Stannigel}, \citenamefont {Zoller},\
  and\ \citenamefont {Kippenberg}}]{24}%
  \BibitemOpen
  \bibfield  {author} {\bibinfo {author} {\bibfnamefont {T.}~\bibnamefont
  {Ramos}}, \bibinfo {author} {\bibfnamefont {V.}~\bibnamefont {Sudhir}},
  \bibinfo {author} {\bibfnamefont {K.}~\bibnamefont {Stannigel}}, \bibinfo
  {author} {\bibfnamefont {P.}~\bibnamefont {Zoller}},\ and\ \bibinfo {author}
  {\bibfnamefont {T.~J.}\ \bibnamefont {Kippenberg}},\ }\bibfield  {title}
  {\bibinfo {title} {Nonlinear quantum optomechanics via individual intrinsic
  two-level defects},\ }\href {https://doi.org/10.1103/PhysRevLett.110.193602}
  {\bibfield  {journal} {\bibinfo  {journal} {Phys. Rev. Lett.}\ }\textbf
  {\bibinfo {volume} {110}},\ \bibinfo {pages} {193602} (\bibinfo {year}
  {2013})}\BibitemShut {NoStop}%
\bibitem [{\citenamefont {Xu}\ \emph {et~al.}(2016)\citenamefont {Xu},
  \citenamefont {Chen},\ and\ \citenamefont {Liu}}]{25}%
  \BibitemOpen
  \bibfield  {author} {\bibinfo {author} {\bibfnamefont {X.-W.}\ \bibnamefont
  {Xu}}, \bibinfo {author} {\bibfnamefont {A.-X.}\ \bibnamefont {Chen}},\ and\
  \bibinfo {author} {\bibfnamefont {Y.-x.}\ \bibnamefont {Liu}},\ }\bibfield
  {title} {\bibinfo {title} {Phonon blockade in a nanomechanical resonator
  resonantly coupled to a qubit},\ }\href
  {https://doi.org/10.1103/PhysRevA.94.063853} {\bibfield  {journal} {\bibinfo
  {journal} {Phys. Rev. A}\ }\textbf {\bibinfo {volume} {94}},\ \bibinfo
  {pages} {063853} (\bibinfo {year} {2016})}\BibitemShut {NoStop}%
\bibitem [{\citenamefont {Wang}\ \emph {et~al.}(2016)\citenamefont {Wang},
  \citenamefont {Miranowicz}, \citenamefont {Li},\ and\ \citenamefont
  {Nori}}]{26}%
  \BibitemOpen
  \bibfield  {author} {\bibinfo {author} {\bibfnamefont {X.}~\bibnamefont
  {Wang}}, \bibinfo {author} {\bibfnamefont {A.}~\bibnamefont {Miranowicz}},
  \bibinfo {author} {\bibfnamefont {H.-R.}\ \bibnamefont {Li}},\ and\ \bibinfo
  {author} {\bibfnamefont {F.}~\bibnamefont {Nori}},\ }\bibfield  {title}
  {\bibinfo {title} {Method for observing robust and tunable phonon blockade in
  a nanomechanical resonator coupled to a charge qubit},\ }\href
  {https://doi.org/10.1103/PhysRevA.93.063861} {\bibfield  {journal} {\bibinfo
  {journal} {Phys. Rev. A}\ }\textbf {\bibinfo {volume} {93}},\ \bibinfo
  {pages} {063861} (\bibinfo {year} {2016})}\BibitemShut {NoStop}%
\bibitem [{\citenamefont {Zhao}\ \emph {et~al.}(2021)\citenamefont {Zhao},
  \citenamefont {Peng}, \citenamefont {Yang}, \citenamefont {Chao},
  \citenamefont {Li},\ and\ \citenamefont {Zhou}}]{27}%
  \BibitemOpen
  \bibfield  {author} {\bibinfo {author} {\bibfnamefont {C.}~\bibnamefont
  {Zhao}}, \bibinfo {author} {\bibfnamefont {R.}~\bibnamefont {Peng}}, \bibinfo
  {author} {\bibfnamefont {Z.}~\bibnamefont {Yang}}, \bibinfo {author}
  {\bibfnamefont {S.}~\bibnamefont {Chao}}, \bibinfo {author} {\bibfnamefont
  {C.}~\bibnamefont {Li}},\ and\ \bibinfo {author} {\bibfnamefont
  {L.}~\bibnamefont {Zhou}},\ }\bibfield  {title} {\bibinfo {title}
  {Atom-mediated phonon blockade and controlled-z gate in superconducting
  circuit system},\ }\href
  {https://doi.org/https://doi.org/10.1002/andp.202100039} {\bibfield
  {journal} {\bibinfo  {journal} {Ann. Phys.}\ }\textbf {\bibinfo {volume}
  {533}},\ \bibinfo {pages} {2100039} (\bibinfo {year} {2021})}\BibitemShut
  {NoStop}%
\bibitem [{\citenamefont {Ohira}\ \emph {et~al.}(2021)\citenamefont {Ohira},
  \citenamefont {Kume}, \citenamefont {Takayama}, \citenamefont {Muralidharan},
  \citenamefont {Takahashi},\ and\ \citenamefont {Toyoda}}]{28}%
  \BibitemOpen
  \bibfield  {author} {\bibinfo {author} {\bibfnamefont {R.}~\bibnamefont
  {Ohira}}, \bibinfo {author} {\bibfnamefont {S.}~\bibnamefont {Kume}},
  \bibinfo {author} {\bibfnamefont {K.}~\bibnamefont {Takayama}}, \bibinfo
  {author} {\bibfnamefont {S.}~\bibnamefont {Muralidharan}}, \bibinfo {author}
  {\bibfnamefont {H.}~\bibnamefont {Takahashi}},\ and\ \bibinfo {author}
  {\bibfnamefont {K.}~\bibnamefont {Toyoda}},\ }\bibfield  {title} {\bibinfo
  {title} {Blockade of phonon hopping in trapped ions in the presence of
  multiple local phonons},\ }\href
  {https://doi.org/10.1103/PhysRevA.103.012612} {\bibfield  {journal} {\bibinfo
   {journal} {Phys. Rev. A}\ }\textbf {\bibinfo {volume} {103}},\ \bibinfo
  {pages} {012612} (\bibinfo {year} {2021})}\BibitemShut {NoStop}%
\bibitem [{\citenamefont {Xie}\ \emph {et~al.}(2017)\citenamefont {Xie},
  \citenamefont {Liao}, \citenamefont {Shang}, \citenamefont {Ye},\ and\
  \citenamefont {Lin}}]{29}%
  \BibitemOpen
  \bibfield  {author} {\bibinfo {author} {\bibfnamefont {H.}~\bibnamefont
  {Xie}}, \bibinfo {author} {\bibfnamefont {C.-G.}\ \bibnamefont {Liao}},
  \bibinfo {author} {\bibfnamefont {X.}~\bibnamefont {Shang}}, \bibinfo
  {author} {\bibfnamefont {M.-Y.}\ \bibnamefont {Ye}},\ and\ \bibinfo {author}
  {\bibfnamefont {X.-M.}\ \bibnamefont {Lin}},\ }\bibfield  {title} {\bibinfo
  {title} {Phonon blockade in a quadratically coupled optomechanical system},\
  }\href {https://doi.org/10.1103/PhysRevA.96.013861} {\bibfield  {journal}
  {\bibinfo  {journal} {Phys. Rev. A}\ }\textbf {\bibinfo {volume} {96}},\
  \bibinfo {pages} {013861} (\bibinfo {year} {2017})}\BibitemShut {NoStop}%
\bibitem [{\citenamefont {Zheng}\ \emph {et~al.}(2019)\citenamefont {Zheng},
  \citenamefont {Yin}, \citenamefont {Bin}, \citenamefont {Lü},\ and\
  \citenamefont {Wu}}]{31}%
  \BibitemOpen
  \bibfield  {author} {\bibinfo {author} {\bibfnamefont {L.-L.}\ \bibnamefont
  {Zheng}}, \bibinfo {author} {\bibfnamefont {T.-S.}\ \bibnamefont {Yin}},
  \bibinfo {author} {\bibfnamefont {Q.}~\bibnamefont {Bin}}, \bibinfo {author}
  {\bibfnamefont {X.-Y.}\ \bibnamefont {Lü}},\ and\ \bibinfo {author}
  {\bibfnamefont {Y.}~\bibnamefont {Wu}},\ }\bibfield  {title} {\bibinfo
  {title} {Single-photon-induced phonon blockade in a hybrid
  spin-optomechanical system},\ }\href
  {https://doi.org/10.1103/PhysRevA.99.013804} {\bibfield  {journal} {\bibinfo
  {journal} {Phys. Rev. A}\ }\textbf {\bibinfo {volume} {99}},\ \bibinfo
  {pages} {013804} (\bibinfo {year} {2019})}\BibitemShut {NoStop}%
\bibitem [{\citenamefont {Sounas}\ and\ \citenamefont {Alu}(2017)}]{36}%
  \BibitemOpen
  \bibfield  {author} {\bibinfo {author} {\bibfnamefont {D.~L.}\ \bibnamefont
  {Sounas}}\ and\ \bibinfo {author} {\bibfnamefont {A.}~\bibnamefont {Alu}},\
  }\bibfield  {title} {\bibinfo {title} {Non-reciprocal photonics based on time
  modulation},\ }\href {https://doi.org/10.1038/s41566-017-0051-x} {\bibfield
  {journal} {\bibinfo  {journal} {Nat. Photon.}\ }\textbf {\bibinfo {volume}
  {11}},\ \bibinfo {pages} {774} (\bibinfo {year} {2017})}\BibitemShut
  {NoStop}%
\bibitem [{\citenamefont {Khanikaev}\ and\ \citenamefont {Shvets}(2017)}]{37}%
  \BibitemOpen
  \bibfield  {author} {\bibinfo {author} {\bibfnamefont {A.~B.}\ \bibnamefont
  {Khanikaev}}\ and\ \bibinfo {author} {\bibfnamefont {G.}~\bibnamefont
  {Shvets}},\ }\bibfield  {title} {\bibinfo {title} {Two-dimensional
  topological photonics},\ }\href {https://doi.org/10.1038/s41566-017-0048-5}
  {\bibfield  {journal} {\bibinfo  {journal} {Nat. Photon.}\ }\textbf {\bibinfo
  {volume} {11}},\ \bibinfo {pages} {763} (\bibinfo {year} {2017})}\BibitemShut
  {NoStop}%
\bibitem [{\citenamefont {Liberal}\ and\ \citenamefont {Engheta}(2017)}]{38}%
  \BibitemOpen
  \bibfield  {author} {\bibinfo {author} {\bibfnamefont {I.}~\bibnamefont
  {Liberal}}\ and\ \bibinfo {author} {\bibfnamefont {N.}~\bibnamefont
  {Engheta}},\ }\bibfield  {title} {\bibinfo {title} {Near-zero refractive
  index photonics},\ }\href {https://doi.org/10.1038/Nphoton.2017.13}
  {\bibfield  {journal} {\bibinfo  {journal} {Nat. Photon.}\ }\textbf {\bibinfo
  {volume} {11}},\ \bibinfo {pages} {149} (\bibinfo {year} {2017})}\BibitemShut
  {NoStop}%
\bibitem [{\citenamefont {Caloz}\ \emph {et~al.}(2018)\citenamefont {Caloz},
  \citenamefont {Alù}, \citenamefont {Tretyakov}, \citenamefont {Sounas},
  \citenamefont {Achouri},\ and\ \citenamefont {Deck-Léger}}]{39}%
  \BibitemOpen
  \bibfield  {author} {\bibinfo {author} {\bibfnamefont {C.}~\bibnamefont
  {Caloz}}, \bibinfo {author} {\bibfnamefont {A.}~\bibnamefont {Alù}},
  \bibinfo {author} {\bibfnamefont {S.}~\bibnamefont {Tretyakov}}, \bibinfo
  {author} {\bibfnamefont {D.}~\bibnamefont {Sounas}}, \bibinfo {author}
  {\bibfnamefont {K.}~\bibnamefont {Achouri}},\ and\ \bibinfo {author}
  {\bibfnamefont {Z.-L.}\ \bibnamefont {Deck-Léger}},\ }\bibfield  {title}
  {\bibinfo {title} {Electromagnetic nonreciprocity},\ }\href
  {https://doi.org/10.1103/PhysRevApplied.10.047001} {\bibfield  {journal}
  {\bibinfo  {journal} {Phys. Rev. Appl.}\ }\textbf {\bibinfo {volume} {10}},\
  \bibinfo {pages} {047001} (\bibinfo {year} {2018})}\BibitemShut {NoStop}%
\bibitem [{\citenamefont {Tokura}\ and\ \citenamefont {Nagaosa}(2018)}]{40}%
  \BibitemOpen
  \bibfield  {author} {\bibinfo {author} {\bibfnamefont {Y.}~\bibnamefont
  {Tokura}}\ and\ \bibinfo {author} {\bibfnamefont {N.}~\bibnamefont
  {Nagaosa}},\ }\bibfield  {title} {\bibinfo {title} {Nonreciprocal responses
  from non-centrosymmetric quantum materials},\ }\href
  {https://doi.org/10.1038/s41467-018-05759-4} {\bibfield  {journal} {\bibinfo
  {journal} {Nat. Commun.}\ }\textbf {\bibinfo {volume} {9}},\ \bibinfo {pages}
  {3740} (\bibinfo {year} {2018})}\BibitemShut {NoStop}%
\bibitem [{\citenamefont {Potton}(2004)}]{41}%
  \BibitemOpen
  \bibfield  {author} {\bibinfo {author} {\bibfnamefont {R.~J.}\ \bibnamefont
  {Potton}},\ }\bibfield  {title} {\bibinfo {title} {Reciprocity in optics},\
  }\href {https://doi.org/10.1088/0034-4885/67/5/R03} {\bibfield  {journal}
  {\bibinfo  {journal} {Rep. Prog. Phys.}\ }\textbf {\bibinfo {volume} {67}},\
  \bibinfo {pages} {717} (\bibinfo {year} {2004})}\BibitemShut {NoStop}%
\bibitem [{\citenamefont {Metelmann}\ and\ \citenamefont {Clerk}(2015)}]{42}%
  \BibitemOpen
  \bibfield  {author} {\bibinfo {author} {\bibfnamefont {A.}~\bibnamefont
  {Metelmann}}\ and\ \bibinfo {author} {\bibfnamefont {A.~A.}\ \bibnamefont
  {Clerk}},\ }\bibfield  {title} {\bibinfo {title} {Nonreciprocal photon
  transmission and amplification via reservoir engineering},\ }\href
  {https://doi.org/10.1103/PhysRevX.5.021025} {\bibfield  {journal} {\bibinfo
  {journal} {Phys. Rev. X}\ }\textbf {\bibinfo {volume} {5}},\ \bibinfo {pages}
  {021025} (\bibinfo {year} {2015})}\BibitemShut {NoStop}%
\bibitem [{\citenamefont {Bernier}\ \emph {et~al.}(2017)\citenamefont
  {Bernier}, \citenamefont {Tóth}, \citenamefont {Koottandavida},
  \citenamefont {Ioannou}, \citenamefont {Malz}, \citenamefont {Nunnenkamp},
  \citenamefont {Feofanov},\ and\ \citenamefont {Kippenberg}}]{43}%
  \BibitemOpen
  \bibfield  {author} {\bibinfo {author} {\bibfnamefont {N.~R.}\ \bibnamefont
  {Bernier}}, \bibinfo {author} {\bibfnamefont {L.~D.}\ \bibnamefont {Tóth}},
  \bibinfo {author} {\bibfnamefont {A.}~\bibnamefont {Koottandavida}}, \bibinfo
  {author} {\bibfnamefont {M.~A.}\ \bibnamefont {Ioannou}}, \bibinfo {author}
  {\bibfnamefont {D.}~\bibnamefont {Malz}}, \bibinfo {author} {\bibfnamefont
  {A.}~\bibnamefont {Nunnenkamp}}, \bibinfo {author} {\bibfnamefont {A.~K.}\
  \bibnamefont {Feofanov}},\ and\ \bibinfo {author} {\bibfnamefont {T.~J.}\
  \bibnamefont {Kippenberg}},\ }\bibfield  {title} {\bibinfo {title}
  {Nonreciprocal reconfigurable microwave optomechanical circuit},\ }\href
  {https://doi.org/10.1038/s41467-017-00447-1} {\bibfield  {journal} {\bibinfo
  {journal} {Nat. Commun.}\ }\textbf {\bibinfo {volume} {8}},\ \bibinfo {pages}
  {604} (\bibinfo {year} {2017})}\BibitemShut {NoStop}%
\bibitem [{\citenamefont {Kamal}\ and\ \citenamefont {Metelmann}(2017)}]{44}%
  \BibitemOpen
  \bibfield  {author} {\bibinfo {author} {\bibfnamefont {A.}~\bibnamefont
  {Kamal}}\ and\ \bibinfo {author} {\bibfnamefont {A.}~\bibnamefont
  {Metelmann}},\ }\bibfield  {title} {\bibinfo {title} {Minimal models for
  nonreciprocal amplification using biharmonic drives},\ }\href
  {https://doi.org/10.1103/PhysRevApplied.7.034031} {\bibfield  {journal}
  {\bibinfo  {journal} {Phys. Rev. Appl.}\ }\textbf {\bibinfo {volume} {7}},\
  \bibinfo {pages} {034031} (\bibinfo {year} {2017})}\BibitemShut {NoStop}%
\bibitem [{\citenamefont {Peterson}\ \emph {et~al.}(2017)\citenamefont
  {Peterson}, \citenamefont {Lecocq}, \citenamefont {Cicak}, \citenamefont
  {Simmonds}, \citenamefont {Aumentado},\ and\ \citenamefont {Teufel}}]{45}%
  \BibitemOpen
  \bibfield  {author} {\bibinfo {author} {\bibfnamefont {G.~A.}\ \bibnamefont
  {Peterson}}, \bibinfo {author} {\bibfnamefont {F.}~\bibnamefont {Lecocq}},
  \bibinfo {author} {\bibfnamefont {K.}~\bibnamefont {Cicak}}, \bibinfo
  {author} {\bibfnamefont {R.~W.}\ \bibnamefont {Simmonds}}, \bibinfo {author}
  {\bibfnamefont {J.}~\bibnamefont {Aumentado}},\ and\ \bibinfo {author}
  {\bibfnamefont {J.~D.}\ \bibnamefont {Teufel}},\ }\bibfield  {title}
  {\bibinfo {title} {Demonstration of efficient nonreciprocity in a microwave
  optomechanical circuit},\ }\href {https://doi.org/10.1103/PhysRevX.7.031001}
  {\bibfield  {journal} {\bibinfo  {journal} {Phys. Rev. X}\ }\textbf {\bibinfo
  {volume} {7}},\ \bibinfo {pages} {031001} (\bibinfo {year}
  {2017})}\BibitemShut {NoStop}%
\bibitem [{\citenamefont {Gu}\ \emph {et~al.}(2017)\citenamefont {Gu},
  \citenamefont {Kockum}, \citenamefont {Miranowicz}, \citenamefont {Liu},\
  and\ \citenamefont {Nori}}]{46}%
  \BibitemOpen
  \bibfield  {author} {\bibinfo {author} {\bibfnamefont {X.}~\bibnamefont
  {Gu}}, \bibinfo {author} {\bibfnamefont {A.~F.}\ \bibnamefont {Kockum}},
  \bibinfo {author} {\bibfnamefont {A.}~\bibnamefont {Miranowicz}}, \bibinfo
  {author} {\bibfnamefont {Y.~X.}\ \bibnamefont {Liu}},\ and\ \bibinfo {author}
  {\bibfnamefont {F.}~\bibnamefont {Nori}},\ }\bibfield  {title} {\bibinfo
  {title} {Microwave photonics with superconducting quantum circuits},\ }\href
  {https://doi.org/10.1016/j.physrep.2017.10.002} {\bibfield  {journal}
  {\bibinfo  {journal} {Phys. Rep.}\ }\textbf {\bibinfo {volume} {718}},\
  \bibinfo {pages} {1} (\bibinfo {year} {2017})}\BibitemShut {NoStop}%
\bibitem [{\citenamefont {Malz}\ \emph {et~al.}(2018)\citenamefont {Malz},
  \citenamefont {Tóth}, \citenamefont {Bernier}, \citenamefont {Feofanov},
  \citenamefont {Kippenberg},\ and\ \citenamefont {Nunnenkamp}}]{47}%
  \BibitemOpen
  \bibfield  {author} {\bibinfo {author} {\bibfnamefont {D.}~\bibnamefont
  {Malz}}, \bibinfo {author} {\bibfnamefont {L.~D.}\ \bibnamefont {Tóth}},
  \bibinfo {author} {\bibfnamefont {N.~R.}\ \bibnamefont {Bernier}}, \bibinfo
  {author} {\bibfnamefont {A.~K.}\ \bibnamefont {Feofanov}}, \bibinfo {author}
  {\bibfnamefont {T.~J.}\ \bibnamefont {Kippenberg}},\ and\ \bibinfo {author}
  {\bibfnamefont {A.}~\bibnamefont {Nunnenkamp}},\ }\bibfield  {title}
  {\bibinfo {title} {Quantum-limited directional amplifiers with
  optomechanics},\ }\href {https://doi.org/10.1103/PhysRevLett.120.023601}
  {\bibfield  {journal} {\bibinfo  {journal} {Phys. Rev. Lett.}\ }\textbf
  {\bibinfo {volume} {120}},\ \bibinfo {pages} {023601} (\bibinfo {year}
  {2018})}\BibitemShut {NoStop}%
\bibitem [{\citenamefont {Shen}\ \emph {et~al.}(2018)\citenamefont {Shen},
  \citenamefont {Zhang}, \citenamefont {Chen}, \citenamefont {Sun},
  \citenamefont {Zou}, \citenamefont {Guo}, \citenamefont {Zou},\ and\
  \citenamefont {Dong}}]{48}%
  \BibitemOpen
  \bibfield  {author} {\bibinfo {author} {\bibfnamefont {Z.}~\bibnamefont
  {Shen}}, \bibinfo {author} {\bibfnamefont {Y.-L.}\ \bibnamefont {Zhang}},
  \bibinfo {author} {\bibfnamefont {Y.}~\bibnamefont {Chen}}, \bibinfo {author}
  {\bibfnamefont {F.-W.}\ \bibnamefont {Sun}}, \bibinfo {author} {\bibfnamefont
  {X.-B.}\ \bibnamefont {Zou}}, \bibinfo {author} {\bibfnamefont {G.-C.}\
  \bibnamefont {Guo}}, \bibinfo {author} {\bibfnamefont {C.-L.}\ \bibnamefont
  {Zou}},\ and\ \bibinfo {author} {\bibfnamefont {C.-H.}\ \bibnamefont
  {Dong}},\ }\bibfield  {title} {\bibinfo {title} {Reconfigurable
  optomechanical circulator and directional amplifier},\ }\href
  {https://doi.org/10.1038/s41467-018-04187-8} {\bibfield  {journal} {\bibinfo
  {journal} {Nat. Commun.}\ }\textbf {\bibinfo {volume} {9}},\ \bibinfo {pages}
  {1797} (\bibinfo {year} {2018})}\BibitemShut {NoStop}%
\bibitem [{\citenamefont {Casimir}(1945)}]{49}%
  \BibitemOpen
  \bibfield  {author} {\bibinfo {author} {\bibfnamefont {H.~B.~G.}\
  \bibnamefont {Casimir}},\ }\bibfield  {title} {\bibinfo {title} {On onsager
  principle of microscopic reversibility},\ }\href
  {https://doi.org/10.1103/RevModPhys.17.343} {\bibfield  {journal} {\bibinfo
  {journal} {Rev. Mod. Phys.}\ }\textbf {\bibinfo {volume} {17}},\ \bibinfo
  {pages} {343} (\bibinfo {year} {1945})}\BibitemShut {NoStop}%
\bibitem [{\citenamefont {Adam}\ \emph {et~al.}(2002)\citenamefont {Adam},
  \citenamefont {Davis}, \citenamefont {Dionne}, \citenamefont {Schloemann},\
  and\ \citenamefont {Stitzer}}]{50}%
  \BibitemOpen
  \bibfield  {author} {\bibinfo {author} {\bibfnamefont {J.~D.}\ \bibnamefont
  {Adam}}, \bibinfo {author} {\bibfnamefont {L.~E.}\ \bibnamefont {Davis}},
  \bibinfo {author} {\bibfnamefont {G.~F.}\ \bibnamefont {Dionne}}, \bibinfo
  {author} {\bibfnamefont {E.~F.}\ \bibnamefont {Schloemann}},\ and\ \bibinfo
  {author} {\bibfnamefont {S.~N.}\ \bibnamefont {Stitzer}},\ }\bibfield
  {title} {\bibinfo {title} {Ferrite devices and materials},\ }\href
  {https://doi.org/10.1109/22.989957} {\bibfield  {journal} {\bibinfo
  {journal} {IEEE Trans. Microw. Theory Tech.}\ }\textbf {\bibinfo {volume}
  {50}},\ \bibinfo {pages} {721} (\bibinfo {year} {2002})}\BibitemShut
  {NoStop}%
\bibitem [{\citenamefont {Dotsch}\ \emph {et~al.}(2005)\citenamefont {Dotsch},
  \citenamefont {Bahlmann}, \citenamefont {Zhuromskyy}, \citenamefont {Hammer},
  \citenamefont {Wilkens}, \citenamefont {Gerhardt}, \citenamefont {Hertel},\
  and\ \citenamefont {Popkov}}]{51}%
  \BibitemOpen
  \bibfield  {author} {\bibinfo {author} {\bibfnamefont {H.}~\bibnamefont
  {Dotsch}}, \bibinfo {author} {\bibfnamefont {N.}~\bibnamefont {Bahlmann}},
  \bibinfo {author} {\bibfnamefont {O.}~\bibnamefont {Zhuromskyy}}, \bibinfo
  {author} {\bibfnamefont {M.}~\bibnamefont {Hammer}}, \bibinfo {author}
  {\bibfnamefont {L.}~\bibnamefont {Wilkens}}, \bibinfo {author} {\bibfnamefont
  {R.}~\bibnamefont {Gerhardt}}, \bibinfo {author} {\bibfnamefont
  {P.}~\bibnamefont {Hertel}},\ and\ \bibinfo {author} {\bibfnamefont {A.~F.}\
  \bibnamefont {Popkov}},\ }\bibfield  {title} {\bibinfo {title} {Applications
  of magneto-optical waveguides in integrated optics: review},\ }\href
  {https://doi.org/10.1364/Josab.22.000240} {\bibfield  {journal} {\bibinfo
  {journal} {J. Opt. Soc. Am. B}\ }\textbf {\bibinfo {volume} {22}},\ \bibinfo
  {pages} {240} (\bibinfo {year} {2005})}\BibitemShut {NoStop}%
\bibitem [{\citenamefont {Shi}\ \emph {et~al.}(2015)\citenamefont {Shi},
  \citenamefont {Yu},\ and\ \citenamefont {Fan}}]{52}%
  \BibitemOpen
  \bibfield  {author} {\bibinfo {author} {\bibfnamefont {Y.}~\bibnamefont
  {Shi}}, \bibinfo {author} {\bibfnamefont {Z.~F.}\ \bibnamefont {Yu}},\ and\
  \bibinfo {author} {\bibfnamefont {S.~H.}\ \bibnamefont {Fan}},\ }\bibfield
  {title} {\bibinfo {title} {Limitations of nonlinear optical isolators due to
  dynamic reciprocity},\ }\href {https://doi.org/10.1038/Nphoton.2015.79}
  {\bibfield  {journal} {\bibinfo  {journal} {Nat. Photon.}\ }\textbf {\bibinfo
  {volume} {9}},\ \bibinfo {pages} {388} (\bibinfo {year} {2015})}\BibitemShut
  {NoStop}%
\bibitem [{\citenamefont {Cao}\ \emph {et~al.}(2017)\citenamefont {Cao},
  \citenamefont {Wang}, \citenamefont {Dong}, \citenamefont {Jing},
  \citenamefont {Liu}, \citenamefont {Chen}, \citenamefont {Ge}, \citenamefont
  {Gong},\ and\ \citenamefont {Xiao}}]{54}%
  \BibitemOpen
  \bibfield  {author} {\bibinfo {author} {\bibfnamefont {Q.-T.}\ \bibnamefont
  {Cao}}, \bibinfo {author} {\bibfnamefont {H.}~\bibnamefont {Wang}}, \bibinfo
  {author} {\bibfnamefont {C.-H.}\ \bibnamefont {Dong}}, \bibinfo {author}
  {\bibfnamefont {H.}~\bibnamefont {Jing}}, \bibinfo {author} {\bibfnamefont
  {R.-S.}\ \bibnamefont {Liu}}, \bibinfo {author} {\bibfnamefont
  {X.}~\bibnamefont {Chen}}, \bibinfo {author} {\bibfnamefont {L.}~\bibnamefont
  {Ge}}, \bibinfo {author} {\bibfnamefont {Q.}~\bibnamefont {Gong}},\ and\
  \bibinfo {author} {\bibfnamefont {Y.-F.}\ \bibnamefont {Xiao}},\ }\bibfield
  {title} {\bibinfo {title} {Experimental demonstration of spontaneous
  chirality in a nonlinear microresonator},\ }\href
  {https://doi.org/10.1103/PhysRevLett.118.033901} {\bibfield  {journal}
  {\bibinfo  {journal} {Phys. Rev. Lett.}\ }\textbf {\bibinfo {volume} {118}},\
  \bibinfo {pages} {033901} (\bibinfo {year} {2017})}\BibitemShut {NoStop}%
\bibitem [{\citenamefont {Bhandare}\ \emph {et~al.}(2005)\citenamefont
  {Bhandare}, \citenamefont {Ibrahim}, \citenamefont {Sandel}, \citenamefont
  {Hongbin}, \citenamefont {Wust},\ and\ \citenamefont {Noe}}]{55}%
  \BibitemOpen
  \bibfield  {author} {\bibinfo {author} {\bibfnamefont {S.}~\bibnamefont
  {Bhandare}}, \bibinfo {author} {\bibfnamefont {S.~K.}\ \bibnamefont
  {Ibrahim}}, \bibinfo {author} {\bibfnamefont {D.}~\bibnamefont {Sandel}},
  \bibinfo {author} {\bibfnamefont {Z.}~\bibnamefont {Hongbin}}, \bibinfo
  {author} {\bibfnamefont {F.}~\bibnamefont {Wust}},\ and\ \bibinfo {author}
  {\bibfnamefont {R.}~\bibnamefont {Noe}},\ }\bibfield  {title} {\bibinfo
  {title} {Novel nonmagnetic 30-db traveling-wave single-sideband optical
  isolator integrated in iii/v material},\ }\href
  {https://doi.org/10.1109/JSTQE.2005.845620} {\bibfield  {journal} {\bibinfo
  {journal} {IEEE J Sel Top Quantum Electron}\ }\textbf {\bibinfo {volume}
  {11}},\ \bibinfo {pages} {417} (\bibinfo {year} {2005})}\BibitemShut
  {NoStop}%
\bibitem [{\citenamefont {Yu}\ and\ \citenamefont
  {Fan}(2009{\natexlab{a}})}]{56}%
  \BibitemOpen
  \bibfield  {author} {\bibinfo {author} {\bibfnamefont {Z.~F.}\ \bibnamefont
  {Yu}}\ and\ \bibinfo {author} {\bibfnamefont {S.~H.}\ \bibnamefont {Fan}},\
  }\bibfield  {title} {\bibinfo {title} {Complete optical isolation created by
  indirect interband photonic transitions},\ }\href
  {https://doi.org/10.1038/Nphoton.2008.273} {\bibfield  {journal} {\bibinfo
  {journal} {Nat. Photon.}\ }\textbf {\bibinfo {volume} {3}},\ \bibinfo {pages}
  {91} (\bibinfo {year} {2009}{\natexlab{a}})}\BibitemShut {NoStop}%
\bibitem [{\citenamefont {Yu}\ and\ \citenamefont
  {Fan}(2009{\natexlab{b}})}]{57}%
  \BibitemOpen
  \bibfield  {author} {\bibinfo {author} {\bibfnamefont {Z.}~\bibnamefont
  {Yu}}\ and\ \bibinfo {author} {\bibfnamefont {S.}~\bibnamefont {Fan}},\
  }\bibfield  {title} {\bibinfo {title} {Optical isolation based on
  nonreciprocal phase shift induced by interband photonic transitions},\ }\href
  {https://doi.org/10.1063/1.3127531} {\bibfield  {journal} {\bibinfo
  {journal} {Appl. Phys. Lett.}\ }\textbf {\bibinfo {volume} {94}},\ \bibinfo
  {pages} {171116} (\bibinfo {year} {2009}{\natexlab{b}})}\BibitemShut
  {NoStop}%
\bibitem [{\citenamefont {Lira}\ \emph {et~al.}(2012)\citenamefont {Lira},
  \citenamefont {Yu}, \citenamefont {Fan},\ and\ \citenamefont {Lipson}}]{58}%
  \BibitemOpen
  \bibfield  {author} {\bibinfo {author} {\bibfnamefont {H.}~\bibnamefont
  {Lira}}, \bibinfo {author} {\bibfnamefont {Z.}~\bibnamefont {Yu}}, \bibinfo
  {author} {\bibfnamefont {S.}~\bibnamefont {Fan}},\ and\ \bibinfo {author}
  {\bibfnamefont {M.}~\bibnamefont {Lipson}},\ }\bibfield  {title} {\bibinfo
  {title} {Electrically driven nonreciprocity induced by interband photonic
  transition on a silicon chip},\ }\href
  {https://doi.org/10.1103/PhysRevLett.109.033901} {\bibfield  {journal}
  {\bibinfo  {journal} {Phys. Rev. Lett.}\ }\textbf {\bibinfo {volume} {109}},\
  \bibinfo {pages} {033901} (\bibinfo {year} {2012})}\BibitemShut {NoStop}%
\bibitem [{\citenamefont {Qin}\ \emph {et~al.}(2014)\citenamefont {Qin},
  \citenamefont {Xu},\ and\ \citenamefont {Wang}}]{59}%
  \BibitemOpen
  \bibfield  {author} {\bibinfo {author} {\bibfnamefont {S.~H.}\ \bibnamefont
  {Qin}}, \bibinfo {author} {\bibfnamefont {Q.}~\bibnamefont {Xu}},\ and\
  \bibinfo {author} {\bibfnamefont {Y.~E.}\ \bibnamefont {Wang}},\ }\bibfield
  {title} {\bibinfo {title} {Nonreciprocal components with distributedly
  modulated capacitors},\ }\href {https://doi.org/10.1109/Tmtt.2014.2347935}
  {\bibfield  {journal} {\bibinfo  {journal} {IEEE Trans. Microw. Theory
  Tech.}\ }\textbf {\bibinfo {volume} {62}},\ \bibinfo {pages} {2260} (\bibinfo
  {year} {2014})}\BibitemShut {NoStop}%
\bibitem [{\citenamefont {Dong}(2015)}]{60}%
  \BibitemOpen
  \bibfield  {author} {\bibinfo {author} {\bibfnamefont {P.}~\bibnamefont
  {Dong}},\ }\bibfield  {title} {\bibinfo {title} {Travelling-wave mach-zehnder
  modulators functioning as optical isolators},\ }\href
  {https://doi.org/10.1364/Oe.23.010498} {\bibfield  {journal} {\bibinfo
  {journal} {Opt. Express}\ }\textbf {\bibinfo {volume} {23}},\ \bibinfo
  {pages} {10498} (\bibinfo {year} {2015})}\BibitemShut {NoStop}%
\bibitem [{\citenamefont {Correas-Serrano}\ \emph {et~al.}(2016)\citenamefont
  {Correas-Serrano}, \citenamefont {Gomez-Diaz}, \citenamefont {Sounas},
  \citenamefont {Hadad}, \citenamefont {Alvarez-Melcon},\ and\ \citenamefont
  {Alu}}]{61}%
  \BibitemOpen
  \bibfield  {author} {\bibinfo {author} {\bibfnamefont {D.}~\bibnamefont
  {Correas-Serrano}}, \bibinfo {author} {\bibfnamefont {J.~S.}\ \bibnamefont
  {Gomez-Diaz}}, \bibinfo {author} {\bibfnamefont {D.~L.}\ \bibnamefont
  {Sounas}}, \bibinfo {author} {\bibfnamefont {Y.}~\bibnamefont {Hadad}},
  \bibinfo {author} {\bibfnamefont {A.}~\bibnamefont {Alvarez-Melcon}},\ and\
  \bibinfo {author} {\bibfnamefont {A.}~\bibnamefont {Alu}},\ }\bibfield
  {title} {\bibinfo {title} {Nonreciprocal graphene devices and antennas based
  on spatiotemporal modulation},\ }\href
  {https://doi.org/10.1109/Lawp.2015.2510818} {\bibfield  {journal} {\bibinfo
  {journal} {IEEE Antennas Wirel. Propag. Lett.}\ }\textbf {\bibinfo {volume}
  {15}},\ \bibinfo {pages} {1529} (\bibinfo {year} {2016})}\BibitemShut
  {NoStop}%
\bibitem [{\citenamefont {Malykin}(2000)}]{62}%
  \BibitemOpen
  \bibfield  {author} {\bibinfo {author} {\bibfnamefont {G.~B.}\ \bibnamefont
  {Malykin}},\ }\bibfield  {title} {\bibinfo {title} {The sagnac effect:
  correct and incorrect explanations},\ }\href
  {https://doi.org/10.1070/pu2000v043n12abeh000830} {\bibfield  {journal}
  {\bibinfo  {journal} {Phys. Usp.}\ }\textbf {\bibinfo {volume} {43}},\
  \bibinfo {pages} {1229} (\bibinfo {year} {2000})}\BibitemShut {NoStop}%
\bibitem [{\citenamefont {Huang}\ \emph {et~al.}(2018)\citenamefont {Huang},
  \citenamefont {Miranowicz}, \citenamefont {Liao}, \citenamefont {Nori},\ and\
  \citenamefont {Jing}}]{63}%
  \BibitemOpen
  \bibfield  {author} {\bibinfo {author} {\bibfnamefont {R.}~\bibnamefont
  {Huang}}, \bibinfo {author} {\bibfnamefont {A.}~\bibnamefont {Miranowicz}},
  \bibinfo {author} {\bibfnamefont {J.-Q.}\ \bibnamefont {Liao}}, \bibinfo
  {author} {\bibfnamefont {F.}~\bibnamefont {Nori}},\ and\ \bibinfo {author}
  {\bibfnamefont {H.}~\bibnamefont {Jing}},\ }\bibfield  {title} {\bibinfo
  {title} {Nonreciprocal photon blockade},\ }\href
  {https://doi.org/10.1103/PhysRevLett.121.153601} {\bibfield  {journal}
  {\bibinfo  {journal} {Phys. Rev. Lett.}\ }\textbf {\bibinfo {volume} {121}},\
  \bibinfo {pages} {153601} (\bibinfo {year} {2018})}\BibitemShut {NoStop}%
\bibitem [{\citenamefont {Li}\ \emph {et~al.}(2019{\natexlab{a}})\citenamefont
  {Li}, \citenamefont {Huang}, \citenamefont {Xu}, \citenamefont {Miranowicz},\
  and\ \citenamefont {Jing}}]{64}%
  \BibitemOpen
  \bibfield  {author} {\bibinfo {author} {\bibfnamefont {B.~J.}\ \bibnamefont
  {Li}}, \bibinfo {author} {\bibfnamefont {R.}~\bibnamefont {Huang}}, \bibinfo
  {author} {\bibfnamefont {X.~W.}\ \bibnamefont {Xu}}, \bibinfo {author}
  {\bibfnamefont {A.}~\bibnamefont {Miranowicz}},\ and\ \bibinfo {author}
  {\bibfnamefont {H.}~\bibnamefont {Jing}},\ }\bibfield  {title} {\bibinfo
  {title} {Nonreciprocal unconventional photon blockade in a spinning
  optomechanical system},\ }\href {https://doi.org/10.1364/Prj.7.000630}
  {\bibfield  {journal} {\bibinfo  {journal} {Photonics Res.}\ }\textbf
  {\bibinfo {volume} {7}},\ \bibinfo {pages} {630} (\bibinfo {year}
  {2019}{\natexlab{a}})}\BibitemShut {NoStop}%
\bibitem [{\citenamefont {Jing}\ \emph {et~al.}(2021)\citenamefont {Jing},
  \citenamefont {Shi},\ and\ \citenamefont {Xu}}]{65}%
  \BibitemOpen
  \bibfield  {author} {\bibinfo {author} {\bibfnamefont {Y.-W.}\ \bibnamefont
  {Jing}}, \bibinfo {author} {\bibfnamefont {H.-Q.}\ \bibnamefont {Shi}},\ and\
  \bibinfo {author} {\bibfnamefont {X.-W.}\ \bibnamefont {Xu}},\ }\bibfield
  {title} {\bibinfo {title} {Nonreciprocal photon blockade and directional
  amplification in a spinning resonator coupled to a two-level atom},\ }\href
  {https://doi.org/10.1103/PhysRevA.104.033707} {\bibfield  {journal} {\bibinfo
   {journal} {Phys. Rev. A}\ }\textbf {\bibinfo {volume} {104}},\ \bibinfo
  {pages} {033707} (\bibinfo {year} {2021})}\BibitemShut {NoStop}%
\bibitem [{\citenamefont {Fang}\ \emph {et~al.}(2012)\citenamefont {Fang},
  \citenamefont {Yu},\ and\ \citenamefont {Fan}}]{66}%
  \BibitemOpen
  \bibfield  {author} {\bibinfo {author} {\bibfnamefont {K.}~\bibnamefont
  {Fang}}, \bibinfo {author} {\bibfnamefont {Z.}~\bibnamefont {Yu}},\ and\
  \bibinfo {author} {\bibfnamefont {S.}~\bibnamefont {Fan}},\ }\bibfield
  {title} {\bibinfo {title} {Photonic aharonov-bohm effect based on dynamic
  modulation},\ }\href {https://doi.org/10.1103/PhysRevLett.108.153901}
  {\bibfield  {journal} {\bibinfo  {journal} {Phys. Rev. Lett.}\ }\textbf
  {\bibinfo {volume} {108}},\ \bibinfo {pages} {153901} (\bibinfo {year}
  {2012})}\BibitemShut {NoStop}%
\bibitem [{\citenamefont {Abdo}\ \emph {et~al.}(2013)\citenamefont {Abdo},
  \citenamefont {Sliwa}, \citenamefont {Frunzio},\ and\ \citenamefont
  {Devoret}}]{67}%
  \BibitemOpen
  \bibfield  {author} {\bibinfo {author} {\bibfnamefont {B.}~\bibnamefont
  {Abdo}}, \bibinfo {author} {\bibfnamefont {K.}~\bibnamefont {Sliwa}},
  \bibinfo {author} {\bibfnamefont {L.}~\bibnamefont {Frunzio}},\ and\ \bibinfo
  {author} {\bibfnamefont {M.}~\bibnamefont {Devoret}},\ }\bibfield  {title}
  {\bibinfo {title} {Directional amplification with a josephson circuit},\
  }\href {https://doi.org/10.1103/PhysRevX.3.031001} {\bibfield  {journal}
  {\bibinfo  {journal} {Phys. Rev. X}\ }\textbf {\bibinfo {volume} {3}},\
  \bibinfo {pages} {031001} (\bibinfo {year} {2013})}\BibitemShut {NoStop}%
\bibitem [{\citenamefont {Tzuang}\ \emph {et~al.}(2014)\citenamefont {Tzuang},
  \citenamefont {Fang}, \citenamefont {Nussenzveig}, \citenamefont {Fan},\ and\
  \citenamefont {Lipson}}]{68}%
  \BibitemOpen
  \bibfield  {author} {\bibinfo {author} {\bibfnamefont {L.~D.}\ \bibnamefont
  {Tzuang}}, \bibinfo {author} {\bibfnamefont {K.}~\bibnamefont {Fang}},
  \bibinfo {author} {\bibfnamefont {P.}~\bibnamefont {Nussenzveig}}, \bibinfo
  {author} {\bibfnamefont {S.~H.}\ \bibnamefont {Fan}},\ and\ \bibinfo {author}
  {\bibfnamefont {M.}~\bibnamefont {Lipson}},\ }\bibfield  {title} {\bibinfo
  {title} {Non-reciprocal phase shift induced by an effective magnetic flux for
  light},\ }\href {https://doi.org/10.1038/Nphoton.2014.177} {\bibfield
  {journal} {\bibinfo  {journal} {Nat. Photon.}\ }\textbf {\bibinfo {volume}
  {8}},\ \bibinfo {pages} {701} (\bibinfo {year} {2014})}\BibitemShut {NoStop}%
\bibitem [{\citenamefont {Sliwa}\ \emph {et~al.}(2015)\citenamefont {Sliwa},
  \citenamefont {Hatridge}, \citenamefont {Narla}, \citenamefont {Shankar},
  \citenamefont {Frunzio}, \citenamefont {Schoelkopf},\ and\ \citenamefont
  {Devoret}}]{69}%
  \BibitemOpen
  \bibfield  {author} {\bibinfo {author} {\bibfnamefont {K.~M.}\ \bibnamefont
  {Sliwa}}, \bibinfo {author} {\bibfnamefont {M.}~\bibnamefont {Hatridge}},
  \bibinfo {author} {\bibfnamefont {A.}~\bibnamefont {Narla}}, \bibinfo
  {author} {\bibfnamefont {S.}~\bibnamefont {Shankar}}, \bibinfo {author}
  {\bibfnamefont {L.}~\bibnamefont {Frunzio}}, \bibinfo {author} {\bibfnamefont
  {R.~J.}\ \bibnamefont {Schoelkopf}},\ and\ \bibinfo {author} {\bibfnamefont
  {M.~H.}\ \bibnamefont {Devoret}},\ }\bibfield  {title} {\bibinfo {title}
  {Reconfigurable josephson circulator/directional amplifier},\ }\href
  {https://doi.org/10.1103/PhysRevX.5.041020} {\bibfield  {journal} {\bibinfo
  {journal} {Phys. Rev. X}\ }\textbf {\bibinfo {volume} {5}},\ \bibinfo {pages}
  {041020} (\bibinfo {year} {2015})}\BibitemShut {NoStop}%
\bibitem [{\citenamefont {Lecocq}\ \emph {et~al.}(2017)\citenamefont {Lecocq},
  \citenamefont {Ranzani}, \citenamefont {Peterson}, \citenamefont {Cicak},
  \citenamefont {Simmonds}, \citenamefont {Teufel},\ and\ \citenamefont
  {Aumentado}}]{70}%
  \BibitemOpen
  \bibfield  {author} {\bibinfo {author} {\bibfnamefont {F.}~\bibnamefont
  {Lecocq}}, \bibinfo {author} {\bibfnamefont {L.}~\bibnamefont {Ranzani}},
  \bibinfo {author} {\bibfnamefont {G.~A.}\ \bibnamefont {Peterson}}, \bibinfo
  {author} {\bibfnamefont {K.}~\bibnamefont {Cicak}}, \bibinfo {author}
  {\bibfnamefont {R.~W.}\ \bibnamefont {Simmonds}}, \bibinfo {author}
  {\bibfnamefont {J.~D.}\ \bibnamefont {Teufel}},\ and\ \bibinfo {author}
  {\bibfnamefont {J.}~\bibnamefont {Aumentado}},\ }\bibfield  {title} {\bibinfo
  {title} {Nonreciprocal microwave signal processing with a field-programmable
  josephson amplifier},\ }\href
  {https://doi.org/10.1103/PhysRevApplied.7.024028} {\bibfield  {journal}
  {\bibinfo  {journal} {Phys. Rev. Appl.}\ }\textbf {\bibinfo {volume} {7}},\
  \bibinfo {pages} {024028} (\bibinfo {year} {2017})}\BibitemShut {NoStop}%
\bibitem [{\citenamefont {Fleury}\ \emph {et~al.}(2014)\citenamefont {Fleury},
  \citenamefont {Sounas}, \citenamefont {Sieck}, \citenamefont {Haberman},\
  and\ \citenamefont {Alu}}]{71}%
  \BibitemOpen
  \bibfield  {author} {\bibinfo {author} {\bibfnamefont {R.}~\bibnamefont
  {Fleury}}, \bibinfo {author} {\bibfnamefont {D.~L.}\ \bibnamefont {Sounas}},
  \bibinfo {author} {\bibfnamefont {C.~F.}\ \bibnamefont {Sieck}}, \bibinfo
  {author} {\bibfnamefont {M.~R.}\ \bibnamefont {Haberman}},\ and\ \bibinfo
  {author} {\bibfnamefont {A.}~\bibnamefont {Alu}},\ }\bibfield  {title}
  {\bibinfo {title} {Sound isolation and giant linear nonreciprocity in a
  compact acoustic circulator},\ }\href
  {https://doi.org/10.1126/science.1246957} {\bibfield  {journal} {\bibinfo
  {journal} {Science}\ }\textbf {\bibinfo {volume} {343}},\ \bibinfo {pages}
  {516} (\bibinfo {year} {2014})}\BibitemShut {NoStop}%
\bibitem [{\citenamefont {Popa}\ and\ \citenamefont {Cummer}(2014)}]{72}%
  \BibitemOpen
  \bibfield  {author} {\bibinfo {author} {\bibfnamefont {B.-I.}\ \bibnamefont
  {Popa}}\ and\ \bibinfo {author} {\bibfnamefont {S.~A.}\ \bibnamefont
  {Cummer}},\ }\bibfield  {title} {\bibinfo {title} {Non-reciprocal and highly
  nonlinear active acoustic metamaterials},\ }\href
  {https://doi.org/10.1038/ncomms4398} {\bibfield  {journal} {\bibinfo
  {journal} {Nat. Commun.}\ }\textbf {\bibinfo {volume} {5}},\ \bibinfo {pages}
  {3398} (\bibinfo {year} {2014})}\BibitemShut {NoStop}%
\bibitem [{\citenamefont {Yang}\ \emph {et~al.}(2015)\citenamefont {Yang},
  \citenamefont {Gao}, \citenamefont {Shi}, \citenamefont {Lin}, \citenamefont
  {Gao}, \citenamefont {Chong},\ and\ \citenamefont {Zhang}}]{73}%
  \BibitemOpen
  \bibfield  {author} {\bibinfo {author} {\bibfnamefont {Z.}~\bibnamefont
  {Yang}}, \bibinfo {author} {\bibfnamefont {F.}~\bibnamefont {Gao}}, \bibinfo
  {author} {\bibfnamefont {X.}~\bibnamefont {Shi}}, \bibinfo {author}
  {\bibfnamefont {X.}~\bibnamefont {Lin}}, \bibinfo {author} {\bibfnamefont
  {Z.}~\bibnamefont {Gao}}, \bibinfo {author} {\bibfnamefont {Y.}~\bibnamefont
  {Chong}},\ and\ \bibinfo {author} {\bibfnamefont {B.}~\bibnamefont {Zhang}},\
  }\bibfield  {title} {\bibinfo {title} {Topological acoustics},\ }\href
  {https://doi.org/10.1103/PhysRevLett.114.114301} {\bibfield  {journal}
  {\bibinfo  {journal} {Phys. Rev. Lett.}\ }\textbf {\bibinfo {volume} {114}},\
  \bibinfo {pages} {114301} (\bibinfo {year} {2015})}\BibitemShut {NoStop}%
\bibitem [{\citenamefont {Zhang}\ \emph {et~al.}(2017)\citenamefont {Zhang},
  \citenamefont {Cheng},\ and\ \citenamefont {Liu}}]{74}%
  \BibitemOpen
  \bibfield  {author} {\bibinfo {author} {\bibfnamefont {T.}~\bibnamefont
  {Zhang}}, \bibinfo {author} {\bibfnamefont {Y.}~\bibnamefont {Cheng}},\ and\
  \bibinfo {author} {\bibfnamefont {X.}~\bibnamefont {Liu}},\ }\bibfield
  {title} {\bibinfo {title} {One-way self-collimated acoustic beams in
  two-dimensional asymmetric sonic crystals with circulating fluids},\ }\href
  {https://doi.org/10.7567/apex.10.067301} {\bibfield  {journal} {\bibinfo
  {journal} {Appl. Phys. Express}\ }\textbf {\bibinfo {volume} {10}},\ \bibinfo
  {pages} {067301} (\bibinfo {year} {2017})}\BibitemShut {NoStop}%
\bibitem [{\citenamefont {Souslov}\ \emph {et~al.}(2017)\citenamefont
  {Souslov}, \citenamefont {van Zuiden}, \citenamefont {Bartolo},\ and\
  \citenamefont {Vitelli}}]{75}%
  \BibitemOpen
  \bibfield  {author} {\bibinfo {author} {\bibfnamefont {A.}~\bibnamefont
  {Souslov}}, \bibinfo {author} {\bibfnamefont {B.~C.}\ \bibnamefont {van
  Zuiden}}, \bibinfo {author} {\bibfnamefont {D.}~\bibnamefont {Bartolo}},\
  and\ \bibinfo {author} {\bibfnamefont {V.}~\bibnamefont {Vitelli}},\
  }\bibfield  {title} {\bibinfo {title} {Topological sound in active-liquid
  metamaterials},\ }\href {https://doi.org/10.1038/Nphys4193} {\bibfield
  {journal} {\bibinfo  {journal} {Nat. Phys.}\ }\textbf {\bibinfo {volume}
  {13}},\ \bibinfo {pages} {1091} (\bibinfo {year} {2017})}\BibitemShut
  {NoStop}%
\bibitem [{\citenamefont {Kim}\ \emph {et~al.}(2017)\citenamefont {Kim},
  \citenamefont {Xu}, \citenamefont {Taylor},\ and\ \citenamefont {Bahl}}]{76}%
  \BibitemOpen
  \bibfield  {author} {\bibinfo {author} {\bibfnamefont {S.}~\bibnamefont
  {Kim}}, \bibinfo {author} {\bibfnamefont {X.}~\bibnamefont {Xu}}, \bibinfo
  {author} {\bibfnamefont {J.~M.}\ \bibnamefont {Taylor}},\ and\ \bibinfo
  {author} {\bibfnamefont {G.}~\bibnamefont {Bahl}},\ }\bibfield  {title}
  {\bibinfo {title} {Dynamically induced robust phonon transport and chiral
  cooling in an optomechanical system},\ }\href
  {https://doi.org/10.1038/s41467-017-00247-7} {\bibfield  {journal} {\bibinfo
  {journal} {Nat. Commun.}\ }\textbf {\bibinfo {volume} {8}},\ \bibinfo {pages}
  {205} (\bibinfo {year} {2017})}\BibitemShut {NoStop}%
\bibitem [{\citenamefont {Barzanjeh}\ \emph {et~al.}(2017)\citenamefont
  {Barzanjeh}, \citenamefont {Wulf}, \citenamefont {Peruzzo}, \citenamefont
  {Kalaee}, \citenamefont {Dieterle}, \citenamefont {Painter},\ and\
  \citenamefont {Fink}}]{77}%
  \BibitemOpen
  \bibfield  {author} {\bibinfo {author} {\bibfnamefont {S.}~\bibnamefont
  {Barzanjeh}}, \bibinfo {author} {\bibfnamefont {M.}~\bibnamefont {Wulf}},
  \bibinfo {author} {\bibfnamefont {M.}~\bibnamefont {Peruzzo}}, \bibinfo
  {author} {\bibfnamefont {M.}~\bibnamefont {Kalaee}}, \bibinfo {author}
  {\bibfnamefont {P.~B.}\ \bibnamefont {Dieterle}}, \bibinfo {author}
  {\bibfnamefont {O.}~\bibnamefont {Painter}},\ and\ \bibinfo {author}
  {\bibfnamefont {J.~M.}\ \bibnamefont {Fink}},\ }\bibfield  {title} {\bibinfo
  {title} {Mechanical on-chip microwave circulator},\ }\href
  {https://doi.org/10.1038/s41467-017-01304-x} {\bibfield  {journal} {\bibinfo
  {journal} {Nat. Commun.}\ }\textbf {\bibinfo {volume} {8}},\ \bibinfo {pages}
  {953} (\bibinfo {year} {2017})}\BibitemShut {NoStop}%
\bibitem [{\citenamefont {Xu}\ \emph {et~al.}(2021)\citenamefont {Xu},
  \citenamefont {Liu}, \citenamefont {Liu},\ and\ \citenamefont {Xiao}}]{78}%
  \BibitemOpen
  \bibfield  {author} {\bibinfo {author} {\bibfnamefont {Y.}~\bibnamefont
  {Xu}}, \bibinfo {author} {\bibfnamefont {J.-Y.}\ \bibnamefont {Liu}},
  \bibinfo {author} {\bibfnamefont {W.}~\bibnamefont {Liu}},\ and\ \bibinfo
  {author} {\bibfnamefont {Y.-F.}\ \bibnamefont {Xiao}},\ }\bibfield  {title}
  {\bibinfo {title} {Nonreciprocal phonon laser in a spinning microwave
  magnomechanical system},\ }\href
  {https://doi.org/10.1103/PhysRevA.103.053501} {\bibfield  {journal} {\bibinfo
   {journal} {Phys. Rev. A}\ }\textbf {\bibinfo {volume} {103}},\ \bibinfo
  {pages} {053501} (\bibinfo {year} {2021})}\BibitemShut {NoStop}%
\bibitem [{\citenamefont {Liu}\ \emph {et~al.}(2019)\citenamefont {Liu},
  \citenamefont {Cai}, \citenamefont {Guo},\ and\ \citenamefont {Yang}}]{79}%
  \BibitemOpen
  \bibfield  {author} {\bibinfo {author} {\bibfnamefont {X.}~\bibnamefont
  {Liu}}, \bibinfo {author} {\bibfnamefont {X.}~\bibnamefont {Cai}}, \bibinfo
  {author} {\bibfnamefont {Q.}~\bibnamefont {Guo}},\ and\ \bibinfo {author}
  {\bibfnamefont {J.}~\bibnamefont {Yang}},\ }\bibfield  {title} {\bibinfo
  {title} {Robust nonreciprocal acoustic propagation in a compact acoustic
  circulator empowered by natural convection},\ }\href
  {https://doi.org/10.1088/1367-2630/ab1bb7} {\bibfield  {journal} {\bibinfo
  {journal} {New J. Phys.}\ }\textbf {\bibinfo {volume} {21}},\ \bibinfo
  {pages} {053001} (\bibinfo {year} {2019})}\BibitemShut {NoStop}%
\bibitem [{\citenamefont {Shen}\ \emph {et~al.}(2019)\citenamefont {Shen},
  \citenamefont {Zhu}, \citenamefont {Li},\ and\ \citenamefont {Cummer}}]{80}%
  \BibitemOpen
  \bibfield  {author} {\bibinfo {author} {\bibfnamefont {C.}~\bibnamefont
  {Shen}}, \bibinfo {author} {\bibfnamefont {X.}~\bibnamefont {Zhu}}, \bibinfo
  {author} {\bibfnamefont {J.}~\bibnamefont {Li}},\ and\ \bibinfo {author}
  {\bibfnamefont {S.~A.}\ \bibnamefont {Cummer}},\ }\bibfield  {title}
  {\bibinfo {title} {Nonreciprocal acoustic transmission in space-time
  modulated coupled resonators},\ }\href
  {https://doi.org/10.1103/PhysRevB.100.054302} {\bibfield  {journal} {\bibinfo
   {journal} {Phys. Rev. B}\ }\textbf {\bibinfo {volume} {100}},\ \bibinfo
  {pages} {054302} (\bibinfo {year} {2019})}\BibitemShut {NoStop}%
\bibitem [{\citenamefont {Sanavio}\ \emph {et~al.}(2020)\citenamefont
  {Sanavio}, \citenamefont {Peano},\ and\ \citenamefont {Xuereb}}]{81}%
  \BibitemOpen
  \bibfield  {author} {\bibinfo {author} {\bibfnamefont {C.}~\bibnamefont
  {Sanavio}}, \bibinfo {author} {\bibfnamefont {V.}~\bibnamefont {Peano}},\
  and\ \bibinfo {author} {\bibfnamefont {A.}~\bibnamefont {Xuereb}},\
  }\bibfield  {title} {\bibinfo {title} {Nonreciprocal topological phononics in
  optomechanical arrays},\ }\href {https://doi.org/10.1103/PhysRevB.101.085108}
  {\bibfield  {journal} {\bibinfo  {journal} {Phys. Rev. B}\ }\textbf {\bibinfo
  {volume} {101}},\ \bibinfo {pages} {085108} (\bibinfo {year}
  {2020})}\BibitemShut {NoStop}%
\bibitem [{\citenamefont {Yang}\ \emph {et~al.}(2020)\citenamefont {Yang},
  \citenamefont {Liu}, \citenamefont {Zhu}, \citenamefont {Liu},\ and\
  \citenamefont {Yang}}]{82}%
  \BibitemOpen
  \bibfield  {author} {\bibinfo {author} {\bibfnamefont {Z.-B.}\ \bibnamefont
  {Yang}}, \bibinfo {author} {\bibfnamefont {J.-S.}\ \bibnamefont {Liu}},
  \bibinfo {author} {\bibfnamefont {A.-D.}\ \bibnamefont {Zhu}}, \bibinfo
  {author} {\bibfnamefont {H.-Y.}\ \bibnamefont {Liu}},\ and\ \bibinfo {author}
  {\bibfnamefont {R.-C.}\ \bibnamefont {Yang}},\ }\bibfield  {title} {\bibinfo
  {title} {Nonreciprocal transmission and nonreciprocal entanglement in a
  spinning microwave magnomechanical system},\ }\href
  {https://doi.org/https://doi.org/10.1002/andp.202000196} {\bibfield
  {journal} {\bibinfo  {journal} {Ann. Phys.}\ }\textbf {\bibinfo {volume}
  {532}},\ \bibinfo {pages} {2000196} (\bibinfo {year} {2020})}\BibitemShut
  {NoStop}%
\bibitem [{\citenamefont {Wang}\ and\ \citenamefont {Lekavicius}(2020)}]{85}%
  \BibitemOpen
  \bibfield  {author} {\bibinfo {author} {\bibfnamefont {H.}~\bibnamefont
  {Wang}}\ and\ \bibinfo {author} {\bibfnamefont {I.}~\bibnamefont
  {Lekavicius}},\ }\bibfield  {title} {\bibinfo {title} {Coupling spins to
  nanomechanical resonators: Toward quantum spin-mechanics},\ }\href
  {https://doi.org/10.1063/5.0024001} {\bibfield  {journal} {\bibinfo
  {journal} {Appl. Phys. Lett.}\ }\textbf {\bibinfo {volume} {117}},\ \bibinfo
  {pages} {230501} (\bibinfo {year} {2020})}\BibitemShut {NoStop}%
\bibitem [{\citenamefont {Rabl}\ \emph {et~al.}(2009)\citenamefont {Rabl},
  \citenamefont {Cappellaro}, \citenamefont {Dutt}, \citenamefont {Jiang},
  \citenamefont {Maze},\ and\ \citenamefont {Lukin}}]{86}%
  \BibitemOpen
  \bibfield  {author} {\bibinfo {author} {\bibfnamefont {P.}~\bibnamefont
  {Rabl}}, \bibinfo {author} {\bibfnamefont {P.}~\bibnamefont {Cappellaro}},
  \bibinfo {author} {\bibfnamefont {M.~V.~G.}\ \bibnamefont {Dutt}}, \bibinfo
  {author} {\bibfnamefont {L.}~\bibnamefont {Jiang}}, \bibinfo {author}
  {\bibfnamefont {J.~R.}\ \bibnamefont {Maze}},\ and\ \bibinfo {author}
  {\bibfnamefont {M.~D.}\ \bibnamefont {Lukin}},\ }\bibfield  {title} {\bibinfo
  {title} {Strong magnetic coupling between an electronic spin qubit and a
  mechanical resonator},\ }\href {https://doi.org/10.1103/PhysRevB.79.041302}
  {\bibfield  {journal} {\bibinfo  {journal} {Phys. Rev. B}\ }\textbf {\bibinfo
  {volume} {79}},\ \bibinfo {pages} {041302} (\bibinfo {year}
  {2009})}\BibitemShut {NoStop}%
\bibitem [{\citenamefont {Kepesidis}\ \emph {et~al.}(2013)\citenamefont
  {Kepesidis}, \citenamefont {Bennett}, \citenamefont {Portolan}, \citenamefont
  {Lukin},\ and\ \citenamefont {Rabl}}]{87}%
  \BibitemOpen
  \bibfield  {author} {\bibinfo {author} {\bibfnamefont {K.~V.}\ \bibnamefont
  {Kepesidis}}, \bibinfo {author} {\bibfnamefont {S.~D.}\ \bibnamefont
  {Bennett}}, \bibinfo {author} {\bibfnamefont {S.}~\bibnamefont {Portolan}},
  \bibinfo {author} {\bibfnamefont {M.~D.}\ \bibnamefont {Lukin}},\ and\
  \bibinfo {author} {\bibfnamefont {P.}~\bibnamefont {Rabl}},\ }\bibfield
  {title} {\bibinfo {title} {Phonon cooling and lasing with nitrogen-vacancy
  centers in diamond},\ }\href {https://doi.org/10.1103/PhysRevB.88.064105}
  {\bibfield  {journal} {\bibinfo  {journal} {Phys. Rev. B}\ }\textbf {\bibinfo
  {volume} {88}},\ \bibinfo {pages} {064105} (\bibinfo {year}
  {2013})}\BibitemShut {NoStop}%
\bibitem [{\citenamefont {Li}\ and\ \citenamefont {Nori}(2018)}]{83}%
  \BibitemOpen
  \bibfield  {author} {\bibinfo {author} {\bibfnamefont {P.-B.}\ \bibnamefont
  {Li}}\ and\ \bibinfo {author} {\bibfnamefont {F.}~\bibnamefont {Nori}},\
  }\bibfield  {title} {\bibinfo {title} {Hybrid quantum system with
  nitrogen-vacancy centers in diamond coupled to surface-phonon polaritons in
  piezomagnetic superlattices},\ }\href
  {https://doi.org/10.1103/PhysRevApplied.10.024011} {\bibfield  {journal}
  {\bibinfo  {journal} {Phys. Rev. Appl.}\ }\textbf {\bibinfo {volume} {10}},\
  \bibinfo {pages} {024011} (\bibinfo {year} {2018})}\BibitemShut {NoStop}%
\bibitem [{\citenamefont {Li}\ \emph {et~al.}(2019{\natexlab{b}})\citenamefont
  {Li}, \citenamefont {Li}, \citenamefont {Zhou}, \citenamefont {Liu},
  \citenamefont {Li},\ and\ \citenamefont {Li}}]{84}%
  \BibitemOpen
  \bibfield  {author} {\bibinfo {author} {\bibfnamefont {B.}~\bibnamefont
  {Li}}, \bibinfo {author} {\bibfnamefont {P.-B.}\ \bibnamefont {Li}}, \bibinfo
  {author} {\bibfnamefont {Y.}~\bibnamefont {Zhou}}, \bibinfo {author}
  {\bibfnamefont {J.}~\bibnamefont {Liu}}, \bibinfo {author} {\bibfnamefont
  {H.-R.}\ \bibnamefont {Li}},\ and\ \bibinfo {author} {\bibfnamefont {F.-L.}\
  \bibnamefont {Li}},\ }\bibfield  {title} {\bibinfo {title} {Interfacing a
  topological qubit with a spin qubit in a hybrid quantum system},\ }\href
  {https://doi.org/10.1103/PhysRevApplied.11.044026} {\bibfield  {journal}
  {\bibinfo  {journal} {Phys. Rev. Appl.}\ }\textbf {\bibinfo {volume} {11}},\
  \bibinfo {pages} {044026} (\bibinfo {year} {2019}{\natexlab{b}})}\BibitemShut
  {NoStop}%
\bibitem [{\citenamefont {Doherty}\ \emph {et~al.}(2013)\citenamefont
  {Doherty}, \citenamefont {Manson}, \citenamefont {Delaney}, \citenamefont
  {Jelezko}, \citenamefont {Wrachtrup},\ and\ \citenamefont {Hollenberg}}]{88}%
  \BibitemOpen
  \bibfield  {author} {\bibinfo {author} {\bibfnamefont {M.~W.}\ \bibnamefont
  {Doherty}}, \bibinfo {author} {\bibfnamefont {N.~B.}\ \bibnamefont {Manson}},
  \bibinfo {author} {\bibfnamefont {P.}~\bibnamefont {Delaney}}, \bibinfo
  {author} {\bibfnamefont {F.}~\bibnamefont {Jelezko}}, \bibinfo {author}
  {\bibfnamefont {J.}~\bibnamefont {Wrachtrup}},\ and\ \bibinfo {author}
  {\bibfnamefont {L.~C.~L.}\ \bibnamefont {Hollenberg}},\ }\bibfield  {title}
  {\bibinfo {title} {The nitrogen-vacancy colour centre in diamond},\ }\href
  {https://doi.org/10.1016/j.physrep.2013.02.001} {\bibfield  {journal}
  {\bibinfo  {journal} {Phys. Rep.}\ }\textbf {\bibinfo {volume} {528}},\
  \bibinfo {pages} {1} (\bibinfo {year} {2013})}\BibitemShut {NoStop}%
\bibitem [{\citenamefont {Bradac}\ \emph {et~al.}(2019)\citenamefont {Bradac},
  \citenamefont {Gao}, \citenamefont {Forneris}, \citenamefont {Trusheim},\
  and\ \citenamefont {Aharonovich}}]{89}%
  \BibitemOpen
  \bibfield  {author} {\bibinfo {author} {\bibfnamefont {C.}~\bibnamefont
  {Bradac}}, \bibinfo {author} {\bibfnamefont {W.}~\bibnamefont {Gao}},
  \bibinfo {author} {\bibfnamefont {J.}~\bibnamefont {Forneris}}, \bibinfo
  {author} {\bibfnamefont {M.~E.}\ \bibnamefont {Trusheim}},\ and\ \bibinfo
  {author} {\bibfnamefont {I.}~\bibnamefont {Aharonovich}},\ }\bibfield
  {title} {\bibinfo {title} {Quantum nanophotonics with group iv defects in
  diamond},\ }\href {https://doi.org/10.1038/s41467-019-13332-w} {\bibfield
  {journal} {\bibinfo  {journal} {Nat. Commun.}\ }\textbf {\bibinfo {volume}
  {10}},\ \bibinfo {pages} {5625} (\bibinfo {year} {2019})}\BibitemShut
  {NoStop}%
\bibitem [{\citenamefont {Li}\ \emph {et~al.}(2020)\citenamefont {Li},
  \citenamefont {Zhou}, \citenamefont {Gao},\ and\ \citenamefont {Nori}}]{90}%
  \BibitemOpen
  \bibfield  {author} {\bibinfo {author} {\bibfnamefont {P.-B.}\ \bibnamefont
  {Li}}, \bibinfo {author} {\bibfnamefont {Y.}~\bibnamefont {Zhou}}, \bibinfo
  {author} {\bibfnamefont {W.-B.}\ \bibnamefont {Gao}},\ and\ \bibinfo {author}
  {\bibfnamefont {F.}~\bibnamefont {Nori}},\ }\bibfield  {title} {\bibinfo
  {title} {Enhancing spin-phonon and spin-spin interactions using linear
  resources in a hybrid quantum system},\ }\href
  {https://doi.org/10.1103/PhysRevLett.125.153602} {\bibfield  {journal}
  {\bibinfo  {journal} {Phys. Rev. Lett.}\ }\textbf {\bibinfo {volume} {125}},\
  \bibinfo {pages} {153602} (\bibinfo {year} {2020})}\BibitemShut {NoStop}%
\bibitem [{\citenamefont {Dong}\ \emph {et~al.}(2021)\citenamefont {Dong},
  \citenamefont {Li}, \citenamefont {Liu},\ and\ \citenamefont {Nori}}]{91}%
  \BibitemOpen
  \bibfield  {author} {\bibinfo {author} {\bibfnamefont {X.-L.}\ \bibnamefont
  {Dong}}, \bibinfo {author} {\bibfnamefont {P.-B.}\ \bibnamefont {Li}},
  \bibinfo {author} {\bibfnamefont {T.}~\bibnamefont {Liu}},\ and\ \bibinfo
  {author} {\bibfnamefont {F.}~\bibnamefont {Nori}},\ }\bibfield  {title}
  {\bibinfo {title} {Unconventional quantum sound-matter interactions in
  spin-optomechanical-crystal hybrid systems},\ }\href
  {https://doi.org/10.1103/PhysRevLett.126.203601} {\bibfield  {journal}
  {\bibinfo  {journal} {Phys. Rev. Lett.}\ }\textbf {\bibinfo {volume} {126}},\
  \bibinfo {pages} {203601} (\bibinfo {year} {2021})}\BibitemShut {NoStop}%
\bibitem [{\citenamefont {Li}\ \emph {et~al.}(2021)\citenamefont {Li},
  \citenamefont {Li}, \citenamefont {Li}, \citenamefont {Gao},\ and\
  \citenamefont {Li}}]{92}%
  \BibitemOpen
  \bibfield  {author} {\bibinfo {author} {\bibfnamefont {X.-X.}\ \bibnamefont
  {Li}}, \bibinfo {author} {\bibfnamefont {P.-B.}\ \bibnamefont {Li}}, \bibinfo
  {author} {\bibfnamefont {H.-R.}\ \bibnamefont {Li}}, \bibinfo {author}
  {\bibfnamefont {H.}~\bibnamefont {Gao}},\ and\ \bibinfo {author}
  {\bibfnamefont {F.-L.}\ \bibnamefont {Li}},\ }\bibfield  {title} {\bibinfo
  {title} {Simulation of topological zak phase in spin-phononic crystal
  networks},\ }\href {https://doi.org/10.1103/PhysRevResearch.3.013025}
  {\bibfield  {journal} {\bibinfo  {journal} {Phys. Rev. Research}\ }\textbf
  {\bibinfo {volume} {3}},\ \bibinfo {pages} {013025} (\bibinfo {year}
  {2021})}\BibitemShut {NoStop}%
\bibitem [{\citenamefont {Shen}\ \emph {et~al.}(2021)\citenamefont {Shen},
  \citenamefont {Dong}, \citenamefont {Chen}, \citenamefont {Qiao},\ and\
  \citenamefont {Li}}]{93}%
  \BibitemOpen
  \bibfield  {author} {\bibinfo {author} {\bibfnamefont {C.-P.}\ \bibnamefont
  {Shen}}, \bibinfo {author} {\bibfnamefont {X.-L.}\ \bibnamefont {Dong}},
  \bibinfo {author} {\bibfnamefont {J.-Q.}\ \bibnamefont {Chen}}, \bibinfo
  {author} {\bibfnamefont {Y.-F.}\ \bibnamefont {Qiao}},\ and\ \bibinfo
  {author} {\bibfnamefont {P.-B.}\ \bibnamefont {Li}},\ }\bibfield  {title}
  {\bibinfo {title} {Strong two-phonon correlations and bound states in the
  continuum in phononic waveguides with embedded siv centers},\ }\href
  {https://doi.org/https://doi.org/10.1002/qute.202100074} {\bibfield
  {journal} {\bibinfo  {journal} {Adv. Quantum Technol.}\ }\textbf {\bibinfo
  {volume} {n/a}},\ \bibinfo {pages} {2100074} (\bibinfo {year}
  {2021})}\BibitemShut {NoStop}%
\bibitem [{\citenamefont {Hepp}\ \emph {et~al.}(2014)\citenamefont {Hepp},
  \citenamefont {Müller}, \citenamefont {Waselowski}, \citenamefont {Becker},
  \citenamefont {Pingault}, \citenamefont {Sternschulte}, \citenamefont
  {Steinmüller-Nethl}, \citenamefont {Gali}, \citenamefont {Maze},
  \citenamefont {Atatüre},\ and\ \citenamefont {Becher}}]{94}%
  \BibitemOpen
  \bibfield  {author} {\bibinfo {author} {\bibfnamefont {C.}~\bibnamefont
  {Hepp}}, \bibinfo {author} {\bibfnamefont {T.}~\bibnamefont {Müller}},
  \bibinfo {author} {\bibfnamefont {V.}~\bibnamefont {Waselowski}}, \bibinfo
  {author} {\bibfnamefont {J.~N.}\ \bibnamefont {Becker}}, \bibinfo {author}
  {\bibfnamefont {B.}~\bibnamefont {Pingault}}, \bibinfo {author}
  {\bibfnamefont {H.}~\bibnamefont {Sternschulte}}, \bibinfo {author}
  {\bibfnamefont {D.}~\bibnamefont {Steinmüller-Nethl}}, \bibinfo {author}
  {\bibfnamefont {A.}~\bibnamefont {Gali}}, \bibinfo {author} {\bibfnamefont
  {J.~R.}\ \bibnamefont {Maze}}, \bibinfo {author} {\bibfnamefont
  {M.}~\bibnamefont {Atatüre}},\ and\ \bibinfo {author} {\bibfnamefont
  {C.}~\bibnamefont {Becher}},\ }\bibfield  {title} {\bibinfo {title}
  {Electronic structure of the silicon vacancy color center in diamond},\
  }\href {https://doi.org/10.1103/PhysRevLett.112.036405} {\bibfield  {journal}
  {\bibinfo  {journal} {Phys. Rev. Lett.}\ }\textbf {\bibinfo {volume} {112}},\
  \bibinfo {pages} {036405} (\bibinfo {year} {2014})}\BibitemShut {NoStop}%
\bibitem [{\citenamefont {Meesala}\ \emph {et~al.}(2018)\citenamefont
  {Meesala}, \citenamefont {Sohn}, \citenamefont {Pingault}, \citenamefont
  {Shao}, \citenamefont {Atikian}, \citenamefont {Holzgrafe}, \citenamefont
  {Gündoğan}, \citenamefont {Stavrakas}, \citenamefont {Sipahigil},
  \citenamefont {Chia}, \citenamefont {Evans}, \citenamefont {Burek},
  \citenamefont {Zhang}, \citenamefont {Wu}, \citenamefont {Pacheco},
  \citenamefont {Abraham}, \citenamefont {Bielejec}, \citenamefont {Lukin},
  \citenamefont {Atatüre},\ and\ \citenamefont {Lončar}}]{95}%
  \BibitemOpen
  \bibfield  {author} {\bibinfo {author} {\bibfnamefont {S.}~\bibnamefont
  {Meesala}}, \bibinfo {author} {\bibfnamefont {Y.-I.}\ \bibnamefont {Sohn}},
  \bibinfo {author} {\bibfnamefont {B.}~\bibnamefont {Pingault}}, \bibinfo
  {author} {\bibfnamefont {L.}~\bibnamefont {Shao}}, \bibinfo {author}
  {\bibfnamefont {H.~A.}\ \bibnamefont {Atikian}}, \bibinfo {author}
  {\bibfnamefont {J.}~\bibnamefont {Holzgrafe}}, \bibinfo {author}
  {\bibfnamefont {M.}~\bibnamefont {Gündoğan}}, \bibinfo {author}
  {\bibfnamefont {C.}~\bibnamefont {Stavrakas}}, \bibinfo {author}
  {\bibfnamefont {A.}~\bibnamefont {Sipahigil}}, \bibinfo {author}
  {\bibfnamefont {C.}~\bibnamefont {Chia}}, \bibinfo {author} {\bibfnamefont
  {R.}~\bibnamefont {Evans}}, \bibinfo {author} {\bibfnamefont {M.~J.}\
  \bibnamefont {Burek}}, \bibinfo {author} {\bibfnamefont {M.}~\bibnamefont
  {Zhang}}, \bibinfo {author} {\bibfnamefont {L.}~\bibnamefont {Wu}}, \bibinfo
  {author} {\bibfnamefont {J.~L.}\ \bibnamefont {Pacheco}}, \bibinfo {author}
  {\bibfnamefont {J.}~\bibnamefont {Abraham}}, \bibinfo {author} {\bibfnamefont
  {E.}~\bibnamefont {Bielejec}}, \bibinfo {author} {\bibfnamefont {M.~D.}\
  \bibnamefont {Lukin}}, \bibinfo {author} {\bibfnamefont {M.}~\bibnamefont
  {Atatüre}},\ and\ \bibinfo {author} {\bibfnamefont {M.}~\bibnamefont
  {Lončar}},\ }\bibfield  {title} {\bibinfo {title} {Strain engineering of the
  silicon-vacancy center in diamond},\ }\href
  {https://doi.org/10.1103/PhysRevB.97.205444} {\bibfield  {journal} {\bibinfo
  {journal} {Phys. Rev. B}\ }\textbf {\bibinfo {volume} {97}},\ \bibinfo
  {pages} {205444} (\bibinfo {year} {2018})}\BibitemShut {NoStop}%
\bibitem [{\citenamefont {Fu}\ \emph {et~al.}(2019)\citenamefont {Fu},
  \citenamefont {Shen}, \citenamefont {Xu}, \citenamefont {Zou}, \citenamefont
  {Cheng}, \citenamefont {Han},\ and\ \citenamefont {Tang}}]{96}%
  \BibitemOpen
  \bibfield  {author} {\bibinfo {author} {\bibfnamefont {W.}~\bibnamefont
  {Fu}}, \bibinfo {author} {\bibfnamefont {Z.}~\bibnamefont {Shen}}, \bibinfo
  {author} {\bibfnamefont {Y.}~\bibnamefont {Xu}}, \bibinfo {author}
  {\bibfnamefont {C.-L.}\ \bibnamefont {Zou}}, \bibinfo {author} {\bibfnamefont
  {R.}~\bibnamefont {Cheng}}, \bibinfo {author} {\bibfnamefont
  {X.}~\bibnamefont {Han}},\ and\ \bibinfo {author} {\bibfnamefont {H.~X.}\
  \bibnamefont {Tang}},\ }\bibfield  {title} {\bibinfo {title} {Phononic
  integrated circuitry and spin–orbit interaction of phonons},\ }\href
  {https://doi.org/10.1038/s41467-019-10852-3} {\bibfield  {journal} {\bibinfo
  {journal} {Nat. Commun.}\ }\textbf {\bibinfo {volume} {10}},\ \bibinfo
  {pages} {2743} (\bibinfo {year} {2019})}\BibitemShut {NoStop}%
\bibitem [{\citenamefont {Liang}\ \emph {et~al.}(2009)\citenamefont {Liang},
  \citenamefont {Yuan},\ and\ \citenamefont {Cheng}}]{97}%
  \BibitemOpen
  \bibfield  {author} {\bibinfo {author} {\bibfnamefont {B.}~\bibnamefont
  {Liang}}, \bibinfo {author} {\bibfnamefont {B.}~\bibnamefont {Yuan}},\ and\
  \bibinfo {author} {\bibfnamefont {J.-c.}\ \bibnamefont {Cheng}},\ }\bibfield
  {title} {\bibinfo {title} {Acoustic diode: Rectification of acoustic energy
  flux in one-dimensional systems},\ }\href
  {https://doi.org/10.1103/PhysRevLett.103.104301} {\bibfield  {journal}
  {\bibinfo  {journal} {Phys. Rev. Lett.}\ }\textbf {\bibinfo {volume} {103}},\
  \bibinfo {pages} {104301} (\bibinfo {year} {2009})}\BibitemShut {NoStop}%
\bibitem [{\citenamefont {Maznev}\ \emph {et~al.}(2013)\citenamefont {Maznev},
  \citenamefont {Every},\ and\ \citenamefont {Wright}}]{98}%
  \BibitemOpen
  \bibfield  {author} {\bibinfo {author} {\bibfnamefont {A.~A.}\ \bibnamefont
  {Maznev}}, \bibinfo {author} {\bibfnamefont {A.~G.}\ \bibnamefont {Every}},\
  and\ \bibinfo {author} {\bibfnamefont {O.~B.}\ \bibnamefont {Wright}},\
  }\bibfield  {title} {\bibinfo {title} {Reciprocity in reflection and
  transmission: What is a `phonon diode'?},\ }\href
  {https://doi.org/10.1016/j.wavemoti.2013.02.006} {\bibfield  {journal}
  {\bibinfo  {journal} {Wave Motion}\ }\textbf {\bibinfo {volume} {50}},\
  \bibinfo {pages} {776} (\bibinfo {year} {2013})}\BibitemShut {NoStop}%
\bibitem [{\citenamefont {Chen}\ \emph {et~al.}(2015)\citenamefont {Chen},
  \citenamefont {Hao}, \citenamefont {Wang}, \citenamefont {Zhang},\ and\
  \citenamefont {Lin}}]{99}%
  \BibitemOpen
  \bibfield  {author} {\bibinfo {author} {\bibfnamefont {S.}~\bibnamefont
  {Chen}}, \bibinfo {author} {\bibfnamefont {C.~C.}\ \bibnamefont {Hao}},
  \bibinfo {author} {\bibfnamefont {C.~H.}\ \bibnamefont {Wang}}, \bibinfo
  {author} {\bibfnamefont {Y.~H.}\ \bibnamefont {Zhang}},\ and\ \bibinfo
  {author} {\bibfnamefont {S.~Y.}\ \bibnamefont {Lin}},\ }\bibfield  {title}
  {\bibinfo {title} {One-dimensional acoustic diodes based on the anisotropy of
  solid media and linear acoustics},\ }\href
  {https://doi.org/10.1016/j.ssc.2015.01.012} {\bibfield  {journal} {\bibinfo
  {journal} {Solid State Commun.}\ }\textbf {\bibinfo {volume} {206}},\
  \bibinfo {pages} {38} (\bibinfo {year} {2015})}\BibitemShut {NoStop}%
\bibitem [{\citenamefont {He}\ and\ \citenamefont {Huang}(2018)}]{100}%
  \BibitemOpen
  \bibfield  {author} {\bibinfo {author} {\bibfnamefont {J.-H.}\ \bibnamefont
  {He}}\ and\ \bibinfo {author} {\bibfnamefont {H.-H.}\ \bibnamefont {Huang}},\
  }\bibfield  {title} {\bibinfo {title} {Multiband switching realized by a
  bidirectionally tunable and multiconfiguration acoustic diode},\ }\href
  {https://doi.org/10.1063/1.5049501} {\bibfield  {journal} {\bibinfo
  {journal} {AIP Adv.}\ }\textbf {\bibinfo {volume} {8}},\ \bibinfo {pages}
  {105032} (\bibinfo {year} {2018})}\BibitemShut {NoStop}%
\bibitem [{\citenamefont {Huang}\ \emph {et~al.}(2020)\citenamefont {Huang},
  \citenamefont {Wang}, \citenamefont {Gong}, \citenamefont {Wu}, \citenamefont
  {Zhang},\ and\ \citenamefont {Zhang}}]{101}%
  \BibitemOpen
  \bibfield  {author} {\bibinfo {author} {\bibfnamefont {Y.}~\bibnamefont
  {Huang}}, \bibinfo {author} {\bibfnamefont {X.}~\bibnamefont {Wang}},
  \bibinfo {author} {\bibfnamefont {X.}~\bibnamefont {Gong}}, \bibinfo {author}
  {\bibfnamefont {H.}~\bibnamefont {Wu}}, \bibinfo {author} {\bibfnamefont
  {D.}~\bibnamefont {Zhang}},\ and\ \bibinfo {author} {\bibfnamefont
  {D.}~\bibnamefont {Zhang}},\ }\bibfield  {title} {\bibinfo {title} {Contact
  nonlinear acoustic diode},\ }\href
  {https://doi.org/10.1038/s41598-020-59270-2} {\bibfield  {journal} {\bibinfo
  {journal} {Sci. Rep.}\ }\textbf {\bibinfo {volume} {10}},\ \bibinfo {pages}
  {2564} (\bibinfo {year} {2020})}\BibitemShut {NoStop}%
\bibitem [{\citenamefont {Bennett}\ and\ \citenamefont
  {DiVincenzo}(2000)}]{102}%
  \BibitemOpen
  \bibfield  {author} {\bibinfo {author} {\bibfnamefont {C.~H.}\ \bibnamefont
  {Bennett}}\ and\ \bibinfo {author} {\bibfnamefont {D.~P.}\ \bibnamefont
  {DiVincenzo}},\ }\bibfield  {title} {\bibinfo {title} {Quantum information
  and computation},\ }\href {https://doi.org/10.1038/35005001} {\bibfield
  {journal} {\bibinfo  {journal} {Nature}\ }\textbf {\bibinfo {volume} {404}},\
  \bibinfo {pages} {247} (\bibinfo {year} {2000})}\BibitemShut {NoStop}%
\bibitem [{\citenamefont {Pauls}\ \emph {et~al.}(2020)\citenamefont {Pauls},
  \citenamefont {Lekavicius},\ and\ \citenamefont {Wang}}]{103}%
  \BibitemOpen
  \bibfield  {author} {\bibinfo {author} {\bibfnamefont {A.}~\bibnamefont
  {Pauls}}, \bibinfo {author} {\bibfnamefont {I.}~\bibnamefont {Lekavicius}},\
  and\ \bibinfo {author} {\bibfnamefont {H.~L.}\ \bibnamefont {Wang}},\
  }\bibfield  {title} {\bibinfo {title} {Coupling silicon vacancy centers in a
  thin diamond membrane to a silica optical microresonator},\ }\href
  {https://doi.org/10.1364/Oe.399331} {\bibfield  {journal} {\bibinfo
  {journal} {Opt. Express}\ }\textbf {\bibinfo {volume} {28}},\ \bibinfo
  {pages} {27300} (\bibinfo {year} {2020})}\BibitemShut {NoStop}%
\bibitem [{\citenamefont {Lang}\ \emph {et~al.}(2020)\citenamefont {Lang},
  \citenamefont {Häußler}, \citenamefont {Fuhrmann}, \citenamefont
  {Waltrich}, \citenamefont {Laddha}, \citenamefont {Scharpf}, \citenamefont
  {Kubanek}, \citenamefont {Naydenov},\ and\ \citenamefont {Jelezko}}]{104}%
  \BibitemOpen
  \bibfield  {author} {\bibinfo {author} {\bibfnamefont {J.}~\bibnamefont
  {Lang}}, \bibinfo {author} {\bibfnamefont {S.}~\bibnamefont {Häußler}},
  \bibinfo {author} {\bibfnamefont {J.}~\bibnamefont {Fuhrmann}}, \bibinfo
  {author} {\bibfnamefont {R.}~\bibnamefont {Waltrich}}, \bibinfo {author}
  {\bibfnamefont {S.}~\bibnamefont {Laddha}}, \bibinfo {author} {\bibfnamefont
  {J.}~\bibnamefont {Scharpf}}, \bibinfo {author} {\bibfnamefont
  {A.}~\bibnamefont {Kubanek}}, \bibinfo {author} {\bibfnamefont
  {B.}~\bibnamefont {Naydenov}},\ and\ \bibinfo {author} {\bibfnamefont
  {F.}~\bibnamefont {Jelezko}},\ }\bibfield  {title} {\bibinfo {title} {Long
  optical coherence times of shallow-implanted, negatively charged silicon
  vacancy centers in diamond},\ }\href {https://doi.org/10.1063/1.5143014}
  {\bibfield  {journal} {\bibinfo  {journal} {Appl. Phys. Lett.}\ }\textbf
  {\bibinfo {volume} {116}},\ \bibinfo {pages} {064001} (\bibinfo {year}
  {2020})}\BibitemShut {NoStop}%
\bibitem [{\citenamefont {Abudi}\ \emph {et~al.}(2021)\citenamefont {Abudi},
  \citenamefont {Douvidzon}, \citenamefont {Bathish},\ and\ \citenamefont
  {Carmon}}]{105}%
  \BibitemOpen
  \bibfield  {author} {\bibinfo {author} {\bibfnamefont {T.~L.}\ \bibnamefont
  {Abudi}}, \bibinfo {author} {\bibfnamefont {M.}~\bibnamefont {Douvidzon}},
  \bibinfo {author} {\bibfnamefont {B.}~\bibnamefont {Bathish}},\ and\ \bibinfo
  {author} {\bibfnamefont {T.}~\bibnamefont {Carmon}},\ }\bibfield  {title}
  {\bibinfo {title} {Resonators made of a disk and a movable
  continuous-membrane},\ }\href {https://doi.org/10.1063/5.0041315} {\bibfield
  {journal} {\bibinfo  {journal} {APL Photonics}\ }\textbf {\bibinfo {volume}
  {6}},\ \bibinfo {pages} {036105} (\bibinfo {year} {2021})}\BibitemShut
  {NoStop}%
\bibitem [{\citenamefont {Zueco}\ \emph {et~al.}(2009)\citenamefont {Zueco},
  \citenamefont {Reuther}, \citenamefont {Kohler},\ and\ \citenamefont
  {Hänggi}}]{106}%
  \BibitemOpen
  \bibfield  {author} {\bibinfo {author} {\bibfnamefont {D.}~\bibnamefont
  {Zueco}}, \bibinfo {author} {\bibfnamefont {G.~M.}\ \bibnamefont {Reuther}},
  \bibinfo {author} {\bibfnamefont {S.}~\bibnamefont {Kohler}},\ and\ \bibinfo
  {author} {\bibfnamefont {P.}~\bibnamefont {Hänggi}},\ }\bibfield  {title}
  {\bibinfo {title} {Qubit-oscillator dynamics in the dispersive regime:
  Analytical theory beyond the rotating-wave approximation},\ }\href
  {https://doi.org/10.1103/PhysRevA.80.033846} {\bibfield  {journal} {\bibinfo
  {journal} {Phys. Rev. A}\ }\textbf {\bibinfo {volume} {80}},\ \bibinfo
  {pages} {033846} (\bibinfo {year} {2009})}\BibitemShut {NoStop}%
\bibitem [{\citenamefont {Matsko}\ \emph {et~al.}(2005)\citenamefont {Matsko},
  \citenamefont {Savchenkov}, \citenamefont {Strekalov},\ and\ \citenamefont
  {Maleki}}]{107}%
  \BibitemOpen
  \bibfield  {author} {\bibinfo {author} {\bibfnamefont {A.~B.}\ \bibnamefont
  {Matsko}}, \bibinfo {author} {\bibfnamefont {A.~A.}\ \bibnamefont
  {Savchenkov}}, \bibinfo {author} {\bibfnamefont {D.}~\bibnamefont
  {Strekalov}},\ and\ \bibinfo {author} {\bibfnamefont {L.}~\bibnamefont
  {Maleki}},\ }\bibfield  {title} {\bibinfo {title} {Whispering gallery
  resonators for studying orbital angular momentum of a photon},\ }\href
  {https://doi.org/10.1103/PhysRevLett.95.143904} {\bibfield  {journal}
  {\bibinfo  {journal} {Phys. Rev. Lett.}\ }\textbf {\bibinfo {volume} {95}},\
  \bibinfo {pages} {143904} (\bibinfo {year} {2005})}\BibitemShut {NoStop}%
\bibitem [{\citenamefont {Lao}()}]{108}%
  \BibitemOpen
  \bibfield  {author} {\bibinfo {author} {\bibfnamefont {B.~Y.}\ \bibnamefont
  {Lao}},\ }\bibfield  {title} {\bibinfo {title} {Gyroscopic effect in surface
  acoustic waves},\ }in\ \href {https://doi.org/10.1109/ULTSYM.1980.197487}
  {\emph {\bibinfo {booktitle} {1980 Ultrasonics Symposium}}},\ p.\ \bibinfo
  {pages} {687}\BibitemShut {NoStop}%
\bibitem [{\citenamefont {Plenio}\ and\ \citenamefont {Knight}(1998)}]{109}%
  \BibitemOpen
  \bibfield  {author} {\bibinfo {author} {\bibfnamefont {M.~B.}\ \bibnamefont
  {Plenio}}\ and\ \bibinfo {author} {\bibfnamefont {P.~L.}\ \bibnamefont
  {Knight}},\ }\bibfield  {title} {\bibinfo {title} {The quantum-jump approach
  to dissipative dynamics in quantum optics},\ }\href
  {https://doi.org/10.1103/RevModPhys.70.101} {\bibfield  {journal} {\bibinfo
  {journal} {Rev. Mod. Phys.}\ }\textbf {\bibinfo {volume} {70}},\ \bibinfo
  {pages} {101} (\bibinfo {year} {1998})}\BibitemShut {NoStop}%
\bibitem [{\citenamefont {Miranowicz}\ \emph {et~al.}(2016)\citenamefont
  {Miranowicz}, \citenamefont {Bajer}, \citenamefont {Lambert}, \citenamefont
  {Liu},\ and\ \citenamefont {Nori}}]{110}%
  \BibitemOpen
  \bibfield  {author} {\bibinfo {author} {\bibfnamefont {A.}~\bibnamefont
  {Miranowicz}}, \bibinfo {author} {\bibfnamefont {J.}~\bibnamefont {Bajer}},
  \bibinfo {author} {\bibfnamefont {N.}~\bibnamefont {Lambert}}, \bibinfo
  {author} {\bibfnamefont {Y.-x.}\ \bibnamefont {Liu}},\ and\ \bibinfo {author}
  {\bibfnamefont {F.}~\bibnamefont {Nori}},\ }\bibfield  {title} {\bibinfo
  {title} {Tunable multiphonon blockade in coupled nanomechanical resonators},\
  }\href {https://doi.org/10.1103/PhysRevA.93.013808} {\bibfield  {journal}
  {\bibinfo  {journal} {Phys. Rev. A}\ }\textbf {\bibinfo {volume} {93}},\
  \bibinfo {pages} {013808} (\bibinfo {year} {2016})}\BibitemShut {NoStop}%
\bibitem [{\citenamefont {Miranowicz}\ \emph {et~al.}(2013)\citenamefont
  {Miranowicz}, \citenamefont {Paprzycka}, \citenamefont {Liu}, \citenamefont
  {Bajer},\ and\ \citenamefont {Nori}}]{111}%
  \BibitemOpen
  \bibfield  {author} {\bibinfo {author} {\bibfnamefont {A.}~\bibnamefont
  {Miranowicz}}, \bibinfo {author} {\bibfnamefont {M.}~\bibnamefont
  {Paprzycka}}, \bibinfo {author} {\bibfnamefont {Y.-x.}\ \bibnamefont {Liu}},
  \bibinfo {author} {\bibfnamefont {J.}~\bibnamefont {Bajer}},\ and\ \bibinfo
  {author} {\bibfnamefont {F.}~\bibnamefont {Nori}},\ }\bibfield  {title}
  {\bibinfo {title} {Two-photon and three-photon blockades in driven nonlinear
  systems},\ }\href {https://doi.org/10.1103/PhysRevA.87.023809} {\bibfield
  {journal} {\bibinfo  {journal} {Phys. Rev. A}\ }\textbf {\bibinfo {volume}
  {87}},\ \bibinfo {pages} {023809} (\bibinfo {year} {2013})}\BibitemShut
  {NoStop}%
\bibitem [{\citenamefont {Hamsen}\ \emph {et~al.}(2017)\citenamefont {Hamsen},
  \citenamefont {Tolazzi}, \citenamefont {Wilk},\ and\ \citenamefont
  {Rempe}}]{112}%
  \BibitemOpen
  \bibfield  {author} {\bibinfo {author} {\bibfnamefont {C.}~\bibnamefont
  {Hamsen}}, \bibinfo {author} {\bibfnamefont {K.~N.}\ \bibnamefont {Tolazzi}},
  \bibinfo {author} {\bibfnamefont {T.}~\bibnamefont {Wilk}},\ and\ \bibinfo
  {author} {\bibfnamefont {G.}~\bibnamefont {Rempe}},\ }\bibfield  {title}
  {\bibinfo {title} {Two-photon blockade in an atom-driven cavity qed system},\
  }\href {https://doi.org/10.1103/PhysRevLett.118.133604} {\bibfield  {journal}
  {\bibinfo  {journal} {Phys. Rev. Lett.}\ }\textbf {\bibinfo {volume} {118}},\
  \bibinfo {pages} {133604} (\bibinfo {year} {2017})}\BibitemShut {NoStop}%
\bibitem [{\citenamefont {Wang}\ \emph
  {et~al.}(2020{\natexlab{c}})\citenamefont {Wang}, \citenamefont {Shen},
  \citenamefont {Zou}, \citenamefont {Fu}, \citenamefont {Shen},\ and\
  \citenamefont {Tang}}]{113}%
  \BibitemOpen
  \bibfield  {author} {\bibinfo {author} {\bibfnamefont {W.}~\bibnamefont
  {Wang}}, \bibinfo {author} {\bibfnamefont {M.}~\bibnamefont {Shen}}, \bibinfo
  {author} {\bibfnamefont {C.-L.}\ \bibnamefont {Zou}}, \bibinfo {author}
  {\bibfnamefont {W.}~\bibnamefont {Fu}}, \bibinfo {author} {\bibfnamefont
  {Z.}~\bibnamefont {Shen}},\ and\ \bibinfo {author} {\bibfnamefont {H.~X.}\
  \bibnamefont {Tang}},\ }\bibfield  {title} {\bibinfo {title}
  {High-acoustic-index-contrast phononic circuits: Numerical modeling},\ }\href
  {https://doi.org/10.1063/5.0019584} {\bibfield  {journal} {\bibinfo
  {journal} {J. Appl. Phys.}\ }\textbf {\bibinfo {volume} {128}},\ \bibinfo
  {pages} {184503} (\bibinfo {year} {2020}{\natexlab{c}})}\BibitemShut
  {NoStop}%
\end{thebibliography}
%

\end{document}